\documentclass[english,prb,aps,twocolumn,floatfix,superscriptaddress]{revtex4-1}

\usepackage[linktocpage,bookmarksopen,bookmarksnumbered]{hyperref}
\usepackage{graphicx}
\usepackage{dcolumn}
\usepackage{amsmath,graphics,epsfig,color,verbatim}
\usepackage{amssymb}
\usepackage{babel}

\begin{document}

\title{Computing total energies in complex materials using charge self-consistent DFT+DMFT
}

\author{Hyowon Park}
\thanks{Present Address: Department of Physics, University of Illinois at Chicago, Chicago, IL 60607, USA}
\email{hyowon@uic.edu}
\affiliation{Department of Applied Physics and Applied Mathematics, Columbia University, New York, NY 10027, USA}
\affiliation{Department of Physics, Columbia University, New York, NY 10027, USA}
\author{Andrew J. Millis}
\affiliation{Department of Physics, Columbia University, New York, NY 10027, USA}
\author{Chris A. Marianetti}
\affiliation{Department of Applied Physics and Applied Mathematics, Columbia University, New York, NY 10027, USA}
%,$^{1,2}$\email{aaa} Andrew J. Millis,$^{2}$ and Chris A. Marianetti$^{1}$}
%
%
%\affiliation{$^{1}$Department of Applied Physics and Applied Mathematics, Columbia University, New York, NY 10027, USA \\
%            $^{2}$Department of Physics, Columbia University, New York, NY 10027, USA}

\date{\today}

\begin{abstract}

We have formulated and implemented a fully charge-self-consistent density functional theory plus dynamical mean field theory methodology which enables an efficient calculation of the total energy of realistic correlated electron systems.  
The density functional portion of the calculation uses a plane wave basis set within the projector augmented wave method
enabling study of systems with large, complex unit cells. The dynamical mean field portion of the calculation is formulated using maximally localized Wannier functions, enabling a convenient  implementation which is independent of the basis set used in the density functional portion of the calculation. The importance of using a correct  double counting term is demonstrated. A generalized form of the standard double counting correction, which we refer to as the $U^\prime$ form,  is described in detail  and used. For comparison the density functional plus U method is implemented within the same framework including the generalized double counting.  The formalism is validated via a calculation of the metal-insulator and structural phase diagrams of the rare-earth nickelate perovskites as functions  of applied pressure and A-site rare-earth ions. The calculated density functional plus dynamical mean field results are found to be  consistent with experiment. The density functional plus U method is shown to grossly overestimate
the tendency for bond-disproportionation and insulating behavior.

\end{abstract}

\maketitle

\section{Introduction}

The combination of density functional theory (DFT) and dynamical mean field theory (DMFT)~\cite{Kotliar:06} has been
successfully applied to the calculation of electronic structures of strongly correlated electronic systems. However, the DFT+DMFT method has mainly been used for the calculation of spectroscopic quantities (especially photoemission) for fixed structures, and the study of energetics and structural properties in complex correlated electron materials remains a formidable challenge within DFT+DMFT.

DFT+DMFT total energy calculations involve significant technical challenges and computational expense, and have been implemented with various degrees of sophistication. Early applications made compromises in the DFT basis set, the definition of the correlated problem solved by DMFT, the method used to solve the DMFT equations, and full charge self-consistency. As the methodology developed these compromises have been removed. An early application  to a realistic material was the computation of the energy versus volume for $\delta$-Pu~\cite{Savrasov:01}. Linear muffin-tin orbitals~\cite{Anderson:75} were used for the DFT basis set and the DMFT equations were solved using a semi-analytic interpolative solver~\cite{Savrasov2005115117}. The volume collapse transition in paramagnetic cerium (Ce) has been studied using numerical Hirsch-Fye quantum Monte Carlo (QMC) calculations to solve the DMFT impurity problem~\cite{Held:01,McMahan:03,Amadon:06}, though the use of the Hirsch-Fye solver required an Ising approximation to the exchange interaction of the impurity problem. More recently, the Jahn-Teller distortions of the wide-gap insulator KCuF$_3$ and of LaMnO$_3$~\cite{Leonov:10} were studied using a plane-wave basis set~\cite{Leonov:08,Leonov:10}, and similar methods were then used to examine the structural transition in paramagnetic iron~\cite{Leonov:11}. However, in these calculations, full charge self-consistency was not attempted. Fully charge self-consistent calculations using the approximate `Hubbard I' impurity solvers~\cite{Hubbard26111963} have been performed to study the elastic properties of Ce~\cite{Pourovskii:07,Amadon:12}, Ce$_2$O$_3$~\cite{Pourovskii:07,Amadon:12}, and Pu$_2$O$_3$~\cite{Amadon:12}. Transition metal systems~\cite{Marco:09} were  studied using a $T$-matrix fluctuation-exchange solver~\cite{Lichtenstein:05}. Very recently, fully charge self-consistent DFT+DMFT calculations using continuous-time QMC~\cite{Werner:06,Werner:2006,Haule:07,Gull_review:11} to solve the DMFT impurity problem, a full-potential linearized augmented plane-wave basis set~\cite{Singh:94}, and projectors to construct the DMFT correlated subspace  have been used to calculate the $z$ position of the As atom in the iron pnictides~\cite{Aichhorn:11,Lee:12}. Calculations of comparable sophistication  were recently executed for the thermodynamics of V$_2$O$_3$~\cite{Lechermann:12} and Ce~\cite{Amadon:13}, but in these calculations a plane-wave basis set within the Projector Augmented Wave (PAW)~\cite{Blochl:1994} framework was used.

Building on this important body of work, we present in this paper a generally applicable and flexible  method for calculating total energies within the DFT+DMFT formalism.  A brief announcement of some of the results has appeared \cite{Park:13}. Similar to Refs.~\onlinecite{Lechermann:12,Amadon:13}, we use a plane-wave basis within the PAW framework, enabling calculations on systems with large and complex unit cells. We define the correlated subspace using  a modified version of maximally localized Wannier functions (MLWF)~\cite{Marzari:97,Marzari:12,Wannier}, which are easily adapted to any basis set used for DFT calculations. The Wannier representation is also  very helpful in performing full charge self-consistency when using a  plane-wave basis because this representation makes it unnecessary  to diagonalize the full plane-wave Hamiltonian at each $k$-point and Matsubara frequency. The DMFT impurity problem is solved using the continuous time QMC method~\cite{Werner:06,Werner:2006,Haule:07,Gull_review:11}.  We draw attention to the importance of the double counting correction and present the details of our $U^\prime$ method that allows control over the magnitude of this term in a manner compatible with full charge self-consistency and the other key aspects of the formalism.

For comparison we also implement the DFT+U method \cite{Anisimov:91} within our formalism by solving  the DMFT impurity problem within Hartree-Fock while keeping all other aspects of the calculation unchanged. This enables a precise understanding of the role of dynamical correlations in complex interacting materials. Obtaining such an understanding has previously been difficult because  most DFT+U implementations  employ an exchange-correlation functional which depends on the spin density (e.g. the local spin-density approximation), while most DFT+DMFT computations, including those of the  present study, utilize a spin-independent exchange-correlation functional (e.g. the local density approximation).  We also show that although DFT+U  provides only a crude approximation to the physics,  the qualitative trends are often useful and the errors across material families can be sufficiently consistent that the method can sometimes serve as a rough proxy for DFT+DMFT.

We demonstrate the power of our methodology by computing the  structural and metal-insulator phase boundaries  of the rare earth nickelate perovskites $R$NiO$_3$ as a function of rare earth ion $R$ and pressure. Additionally, we provide bond-length differences as a function of pressure for numerous rare-earth ions, and compute total energy as a function of bond disproportionation for different pressures. 
These calculations provide a critical test of the DFT+DMFT method because they require resolving  small energy differences between subtly different  structures in a situation where  standard DFT calculations fail. Further they require a method which is accurate for both metallic and insulating phases. We show that the DFT+U approach grossly overestimate the tendency to order, while our fully charge self-consistent DFT+DMFT calculations accurately capture the physics in this system.

This paper is organized as follows. In Sec.$\:$\ref{sec:Theory-0} and Sec.$\:$\ref{sec:Theory-1}, we present the formalism of our DFT+DMFT method with particular attention to the issues arising when using the MLWF orbitals to define the correlated subspace.  We then derive the formula to compute the charge density  within our DFT+DMFT implementation in Sec.$\:$\ref{sec:Theory-2}. The full implementation of the charge self-consistent calculation is given in Sec.$\:$\ref{sec:Theory-3}. The total energy formula is derived in Sec.$\:$\ref{sec:Theory-4} and we present the double counting formula used throughout this paper in Sec.$\:$\ref{sec:Theory-5}. 
In Sec.$\:$\ref{sec:Results}, we apply our DFT+DMFT method to the $ab$-$initio$ calculation of rare-earth nickelates. We first overview the structural and electronic properties of the rare-earth nickelates (Sec.$\:$\ref{sec:Results-1}) and explain the aspects of computing the phase transition in these materials by displaying the total energy and the many-body density of states at the Fermi level as a function of $\delta a$ at a fixed pressure (Sec.$\:$\ref{sec:Results-2}). We then show the main results of the structural and metal-insulator transition phase diagram of rare-earth nickelates as functions of pressure and rare-earth ions (Sec.$\:$\ref{sec:Results-3}) in addition to the Ni-O bond-length disproportionation $\delta a$ results as a function of pressure (Sec.$\:$\ref{sec:Results-4}) obtained from our DFT+DMFT total energy calculations and compare the results to experiment and to DFT+U.
In Sec.$\:$\ref{sec:Result_dc-1}, Sec.$\:$\ref{sec:Result_dc-2}, and Sec.$\:$\ref{sec:Result_dc-3},  we explain the effect of the double counting  on the phase diagram and show that the particular form of the double counting used here is  physically reasonable as is demonstrated by a comparison of the DMFT spectral function to experimental spectra.

\section{DFT+DMFT implementation}

\label{sec:Theory}

In this section, we present the specifics of our  implementation of the DFT+DMFT formalism. This is a `beyond DFT' methodology in which a subset of the electronic degrees of freedom (``the correlated subspace'') are treated by a sophisticated many-body physics method while the remaining degrees of freedom are treated within density functional theory (we use the generalized gradient approximation in a plane-wave basis in conjunction with the PAW formalism~\cite{Blochl:1994}). The crucial issues in any beyond-DFT methodologies are the construction of the correlated subspace (we use maximally localized Wannier functions), the method of solving the correlation problem (we use the single-site dynamical mean field approximation), and the embedding of the correlated subspace into the wider electronic structure (key issues are full charge self-consistency and  the double-counting correction, both discussed in details below).

We begin by recapitulating the general theory, to establish notation and highlight the aspects important for our subsequent discussion. We then discuss in detail the definition of the correlated subspace and conclude this section by presenting the full self-consistency loop, along with a discussion of the issues that arise in practical implementations.

\subsection{DFT+DMFT: General theory}

\label{sec:Theory-0}

The DFT+DMFT method can be formally defined ~\cite{Savrasov:04,Kotliar:06} in terms of a functional $\Gamma$ of four variables: the total charge density $\rho$,  the local Green's function $G_{cor}$ associated with a correlated subspace which is treated with a beyond-DFT method, an effective potential  $V^{Hxc}$ conjugate to a charge density,  and a local self energy $\Sigma_{cor}$ conjugate to $G_{cor}$:
\begin{eqnarray}
\Gamma[\rho,G_{cor}; V^{Hxc},\Sigma] &=& \mbox{Tr}\left[\mbox{ln} G\right]+\Phi[\rho,G_{cor}] \\ \nonumber
&&-\mbox{Tr}[V^{Hxc}\rho]-\mbox{Tr}[\Sigma_{cor}G_{cor}]
\label{eq:SDFT}
\end{eqnarray}
Here $G$ is a Green's function defined in the continuum as follows:
\begin{equation}
G=\left(i\omega_n+\mu+\frac{1}{2}\nabla^2-V^{ext}-V^{Hxc}-P_{cor}^{\dagger}\Sigma_{cor}P_{cor}\right)^{-1}
\label{Gdef}
\end{equation}
where $\mu$ is the chemical potential and $V^{ext}$ is a potential arising from the ions and any externally applied fields. $G_{cor}$ and $\Sigma_{cor}$ are operators acting in the continuum but with non-zero matrix elements only in the correlated subspace.
$P_{cor}$ ($P^\dagger_{cor}$)  is a projection operator defined to downfold
(upfold) between the correlated subpace and the space in which $G$ is
defined. For example, if $G$ is defined in the position representation and
the correlated subspace is spanned by a set of states $\{ |\phi_i\rangle \}$, then
$P_{cor}=\int dx \sum_{i} | \phi_i \rangle \langle \phi_i |x\rangle \langle x|$.
% and $P_{cor}$ is an operator which is used to upfold from the correlated subspace to the full basis (in the present case, the position representation). 
It should be noted that $i\omega_n-G^{-1}$ is a frequency dependent, non-hermitian operator that plays the role of an effective Hamiltonian analogous to the Kohn-Sham Hamiltonian in DFT.

$\Phi$ encodes the functional dependence of the free energy arising from electron-electron interactions.   If one omitted the variables $G_{cor}$ and $\Sigma_{cor}$, then $\Phi$ would be the universal Hohenberg-Kohn functional familiar from density functional theory. Alternatively, if  $\rho$ and $V^{Hxc}$ are omitted and $P_{cor}=1$,  then $\Phi$ would be the Luttinger-Ward functional defined from all vacuum to vacuum diagrams with appropriate symmetry factors.

Demanding that $\Gamma$ be stationary with respect to variations of $\rho$, $G_{cor}$, $V^{Hxc}$, and $\Sigma_{cor}$ yields
\begin{eqnarray}\label{VHxcdef} 
V^{Hxc}&=\frac{\delta \Phi[\rho,G_{cor}] }{\delta \rho} 
\\
 \Sigma_{cor}&=\frac{\delta \Phi[\rho,G_{cor}] }{\delta G_{cor}} 
 \label{Sigmadef}
 \\
 \rho&=\mbox{Tr}~G
 \label{rhodef}
 \\
 G_{cor}&=P_{cor}GP^\dagger_{cor}
\label{SCE}
\end{eqnarray}

The equations above provide a formal specification of the theory. To proceed we need to introduce approximations. In the DFT+DMFT methodology $\Phi[\rho,G_{cor}]$  is \emph{approximated} as follows:
\begin{align}\label{}
\Phi[\rho,G_{cor}]\approx \Phi_\rho[\rho] + \Phi_G[G_{cor}]
\end{align}
where  $\Phi_\rho[\rho]$ is the universal functional of density functional theory and has no explicit dependence on $G_{cor}$ and $\Phi_G[G_{cor}]$ is the Luttinger-Ward functional of the model describing the correlated states and has no explicit dependence on $\rho$.  Implicit in the construction of $\Phi_G[G_{cor}]$ is a specification of interactions that couple the degrees of freedom in the correlated subspace.

The sum $\Phi_\rho[\rho]+\Phi_G[G_{cor}]$ must then be  corrected by subtracting a ``double counting'' term that removes the terms which depend on the density in the correlated subspace and are included in both $\Phi_\rho$ and $\Phi_G$, thus:
\begin{equation}
\Phi[\rho,G_{cor}]\approx\Phi_\rho[\rho]+\Phi_G[G_{cor}]-\Phi_{DC}[\rho_{cor}]
\label{PhiDFT+DMFT}
\end{equation}
where $\rho_{cor}$ is the total density in the correlated subspace.
Proceeding further, we take $\Phi_\rho$ to be the sum of the  Hartree term  and  the Perdue-Burke-Ernzerhof generalized gradient approximation (GGA)  approximation to the exchange-correlation functional ~\cite{Perdew:96}:
\begin{equation}
\Phi_\rho[\rho]\rightarrow  \frac{1}{2} \int d\mathbf{r} \int d\mathbf{r}' \frac{\rho(\mathbf{r})\rho(\mathbf{r}')}{|\mathbf{r}-\mathbf{r}'|} 
+E_{xc}^{GGA}[\rho]
\label{GGA}
\end{equation}
We further treat the correlated subspace within the single-site dynamical mean field approximation so that the only important part of the correlated Green's function is the onsite (local) Green's function $G_{loc}$ and the double counting correction depends on the occupancy $N_d$ computed from the local Green's function of the correlated orbitals; thus
\begin{equation}
\Phi_{G}-\Phi_{DC}\rightarrow\Phi^{DMFT}[G_{loc}]-E^{DC}[N_d]
\label{DMFT+DC}
\end{equation}
Correspondingly, $\Sigma_{cor}$ is $\Sigma_{loc}-V^{DC}$ where $\Sigma_{loc}$ is $\delta \Phi^{DMFT}/ \delta G_{loc}$ obtained from the solution of the dynamical mean field equations and $V^{DC}$ is $\delta E^{DC}/ \delta N_d$.
$V^{Hxc}$ is the functional derivative of $\Phi_\rho[\rho]$ with respect to $\rho$.

\subsection{Construction of the correlated subspace and the hybridization window}

\label{sec:Theory-1}

% Technical part

Implementation of the formalism described above requires a prescription for the correlated subspace. It is also useful to define the  ``hybridization window", which refers to the range of states which hybridize with the correlated subspace. The hybridization window plays an important role in our  Wannier function-based construction of the correlated subspace.

Our choice of the correlated subspace is guided by the use of the GGA and DMFT to perform calculations.  Given that DMFT is optimized for recovering local physics, it seems reasonable to construct the correlated subspace from  local orbitals which  most accurately represent the states in which correlations are strong.  To define these states we use  the Marzari-Vanderbilt Maximally Localized Wannier Function (MLWF) procedure \cite{Marzari:97}, which constructs localized states as appropriately phased linear combinations of band states within an energy window. In our formalism, this energy window used in the MLWF procedure is, by construction, the hybridization window. We choose the energy window to be wide enough that the correlated subspace (i.e. a subset of the Wannier functions) are sufficiently localized and resemble the atomic states of interest (i.e. $d$-like orbitals, in the study of transition metal oxides).

The Wannier representation has an added advantage.  The presence of the self energy operator means that a straightforward  computation of $G$ (Eq.~\ref{Gdef}) in a large basis (e.g. plane waves) is cumbersome, requiring that one diagonalize the operator at every basis state (here, $\mathbf{k}$-point) and at every Matsubara frequency.  While massive parallelization can mitigate the problem, it is advantageous to circumvent the issue.  The complete basis can be decomposed into a block composed of all Wannier functions (ie. all states in the hybridization window) and another block consisting of all remaining states. By construction both the full and the bare Green's function are block diagonal, with one block having matrix elements only among states within the hybridization window and the other  having only matrix elements between states not in the hybridization window.  Thus the matrix inversion required to construct the non-trivial part of the Green's function can always be performed in a compact representation.

The MLWF $|W_n^{\mathbf{R}}\rangle$ are labeled with a vector $\mathbf{R}$ indicating the unit cell and a two-part index $n=(\tau,\alpha)$ in which $\tau$ labels an atom at relative position $\mathbf{R}_{\tau}$ in the unit cell and $\alpha$ labels the orbital character referenced to the corresponding site.
The MLWF  are defined as a 
linear combination of the Kohn-Sham (KS) wavefunctions $\psi_{i\mathbf{k}}$ in a given energy range:
\begin{eqnarray}
| W_n^{\mathbf{R}} \rangle & = & \frac{1}{\sqrt{N_{\mathbf{k}}}}\sum_{\mathbf{k},i}
e^{-i\mathbf{k}\cdot\mathbf{R}}U_{ni}^{\mathbf{k}}|\psi_{i\mathbf{k}} \rangle
\label{eq:Wan}
\end{eqnarray}
and will normally be centered at position $\mathbf{R}+\mathbf{R}_{\tau}$.
The unitary matrices $U_{ni}^{\mathbf{k}}$ are chosen to minimize a spread functional~\cite{Marzari:97}. 
The band index $i$ runs over an energy range that defined by the hybridization window. By construction, correlated orbitals defined in terms of Wannier functions cannot mix with states outside of the hybridization window.

After computing the $|W_n^{\mathbf{R}}\rangle$ we perform an additional unitary transform $\hat{\Lambda}$ representing the rotations of orbitals in the correlated subspace in order to minimize the off-diagonal matrix elements within each site-sector of the local correlated manifold and hence minimize the off-diagonal components of  $\Sigma_{loc}$. This transformation is very useful in practice since quantum impurity models with  diagonal or nearly diagonal hybridization matrices can be much more efficiently solved numerically~\cite{Gull_review:11}.  
The details of computing $\hat{\Lambda}$ are explained in Appendix A.

The final unitary transform  from the KS wavefunction to the Wannier basis is thus given by
\begin{equation}
\bar{U}^{\mathbf{k}}_{mi} = \sum_n \Lambda_{mn}\cdot U^{\mathbf{k}}_{ni} 
\label{eq:WanU}
\end{equation}
where $\hat{\Lambda}$ is the unitary matrix satisfying the minimization of off-diagnoal matrix elements of the correlated Hamiltonian (see Eq.$\:$\ref{eq:rot}).
Therefore, the rotated Wannier function $\bar{W}$ is defined by
\begin{eqnarray}
| \bar{W}_m^{\mathbf{R}}\rangle & = & \frac{1}{\sqrt{N_{\mathbf{k}}}}\sum_{\mathbf{k},i}
e^{-i\mathbf{k}\cdot\mathbf{R}}\bar{U}_{mi}^{\mathbf{k}}\psi_{i\mathbf{k}}(\mathbf{r})
\label{eq:Wan2}
\end{eqnarray}

\subsection{Charge density in DFT+DMFT} 

\label{sec:Theory-2}

An important step in the  full implementation of our DFT+DMFT method is the construction of the full charge density. Modern plane-wave codes use either an  ultra-soft pseudo (PS) potential or a PAW formalism. In this formalism there are two contributions to the local charge density: from the PS wavefunctions $\tilde{\Psi}$  and from an ``augmentation charge term'' expressing the difference between the PS wavefunctions and the KS wave functions $\psi^{KS}$ corresponding to the full potential. In our approach the soft and augmentation charge must be expressed in the Wannier representation which is convenient for calculation of the Green's function in the correlated energy window. In this subsection we present the needed formalism. The resulting methodology is similar to the  charge self-consistent PAW+DMFT scheme derived  for the projected local orbital  basis set in Refs.~\onlinecite{Lechermann:06,Amadon:12}.

The fundamental definition of the  charge density $\rho$ is  from the Green's function via Eq.~\ref{rhodef}. Expressing $G$ in the band (ij) basis we have
\begin{equation}
\rho(\mathbf{r})=\frac{1}{N_{\mathbf{k}}}\sum_{\mathbf{k};ij}n_{\mathbf{k};ij}\left<\psi^{KS}_{\mathbf{k}i}|\mathbf{r}\right>\left<\mathbf{r}|\psi^{KS}_{\mathbf{k}j}\right>
\label{rhodef1}
\end{equation}
where the density matrix in the band basis is
\begin{eqnarray}
n_{\mathbf{k};ij} & = & T\sum_{i\omega_n}G_{\mathbf{k};ij}(i\omega_n)e^{i\omega_n\cdot 0^-}.
\label{eq:eq1_1}
\end{eqnarray}
and $T$ is the temperature.

We observe that for states outside the hybridization window, $W$, $G=G^0$ so the density matrix $n_{\mathbf{k};ij}$ becomes the Fermi function $f_{\mathbf{k}i}\delta_{ij}$ with $f_{\mathbf{k}i}$ being the Fermi function for state $\mathbf{k}$ in band $i$. Alternatively, for the bands within the hybridization window the density matrix is most easily computed from the  Wannier $(mn)$ representation as 
\begin{eqnarray}
n_{\mathbf{k}ij} & = & T\sum_{i\omega_n}e^{i\omega_n\cdot 0^-}\sum_{mn} \bar{U}^{\mathbf{k}}_{mi}G_{\mathbf{k}mn}(i\omega_n)\bar{U}^{\mathbf{k} *}_{nj}.
\label{eq:eq2}
\end{eqnarray}
so that
\begin{equation}
\rho(\mathbf{r})  =  \sum_{i\notin W}\rho^{DFT}_i(\mathbf{r}) + \sum_{i,j\in W}\rho^{DMFT}_{ij}(\mathbf{r}) 
\label{eq:eq3}
\end{equation}
with 
\begin{equation}
\rho^{DMFT}_{ij}(\mathbf{r})=\frac{1}{N_{\mathbf{k}}}\sum_{\mathbf{k}}
n_{\mathbf{k}ij} \langle\psi^{KS}_{\mathbf{k}i}|\mathbf{r}\rangle\langle\mathbf{r}|\psi^{KS}_{\mathbf{k}j}\rangle
\label{rhoDMFT}
\end{equation} 
and
\begin{equation}
\rho^{DFT}_i(\mathbf{r})=\frac{1}{N_{\mathbf{k}}}\sum_{\mathbf{k}}f_{\mathbf{k}i}\left<\psi^{KS}_{\mathbf{k}i}|\mathbf{r}\right>\left<\mathbf{r}|\psi^{KS}_{\mathbf{k}i}\right>
\label{rhoDFTi}
\end{equation}

Within the PAW formalism, the KS wavefunction $\psi^{KS}_{\mathbf{k}i}$ is related to  the PS wavefunction $\tilde{\psi}_{\mathbf{k}i}$ by a linear transformation $\hat{T}$, i.e, $\psi^{KS}=\hat{T}\tilde{\psi}$. An operator $\hat{O}$ acting on $\tilde{\psi}$ transforms as $\hat{T}^{\dagger}\hat{O}\hat{T}$. As a result, the charge density $\rho$ can be split into the soft-charge-density term $\tilde{\rho}$, the on-site all-electron charge-density term $\rho^1$, and the on-site PS charge-density term $\tilde{\rho}^1$, i.e.,
\begin{eqnarray}
\hat{T}^{\dagger}\rho\hat{T} & = & \tilde{\rho}+\rho^1-\tilde{\rho}^1.
\label{eq:eq5}
\end{eqnarray}
The calculation of these terms within DFT is explained in Ref.$\:$\onlinecite{Kresse19991758}.

As explained above, the charge density within DMFT is computed using the DMFT density matrix $n_{\mathbf{k}ij}$
instead of the Fermi occupancy $f_{\mathbf{k}i}$ within DFT. As a result, the soft charge $\tilde{\rho}$ is given by
\begin{eqnarray}
\tilde{\rho}^{DMFT}(\mathbf{r}) & = & 
\frac{1}{N_{\mathbf{k}}}\sum_{i,j,\mathbf{k}}
n_{\mathbf{k}ij} \langle\tilde{\psi}_{\mathbf{k}i}|\mathbf{r}\rangle\langle\mathbf{r}|\tilde{\psi}_{\mathbf{k}j}\rangle.
\label{eq:eq6}
\end{eqnarray}
The on-site charges are given by the usual PAW formula
\begin{eqnarray}
\rho^{1,DMFT}(\mathbf{r}) & = & \sum_{m,n}\bar{\rho}_{mn}\cdot\langle\phi_m|\mathbf{r}\rangle\langle\mathbf{r}|\phi_n\rangle \label{eq:eq7}\\
\tilde{\rho}^{1,DMFT}(\mathbf{r}) & = & \sum_{m,n}\bar{\rho}_{mn}\cdot\langle\tilde{\phi}_m|\mathbf{r}\rangle\langle\mathbf{r}|\tilde{\phi}_n\rangle.
\label{eq:eq7_2}
\end{eqnarray}
where $|\phi_n\rangle$ are the all-electron partial waves and  $|\tilde{\phi}_n\rangle$ are the PS partial waves.
Here, the occupancy $\bar{\rho}_{mn}$ of an augmentation channel $(m,n)$ is given by
\begin{equation}
\bar{\rho}_{mn} = \frac{1}{N_{\mathbf{k}}}\sum_{\mathbf{k}ij}\langle\tilde{p}_n|\tilde{\psi}_{\mathbf{k}j}\rangle\cdot n_{\mathbf{k}ij}\cdot\langle\tilde{\psi}_{\mathbf{k}i}|\tilde{p}_m\rangle
\label{eq:eq8}
\end{equation}
where $|\tilde{p}_n\rangle$ are the projector functions which are dual to the PS partial waves.

The sum over band indices $i,j$ in Eq.$\:$\ref{eq:eq6} and Eq.$\:$\ref{eq:eq8} can be simplified to the sum over one index because the density matrix  $n_{ij}$ is Hermitian and so can be written in terms of eigenvalues $w_{\mathbf{k}\lambda}$ and eigenfunctions $\phi_\lambda$ as 
\begin{equation}
n_{\mathbf{k}ij} = \sum_{\lambda}U_{\mathbf{k}i\lambda}^{DMFT}\cdot 
w_{\mathbf{k}\lambda}\cdot U_{\mathbf{k}j\lambda}^{DMFT*}
\end{equation}
where $U_{\mathbf{k}}^{DMFT}$ are unitary matrices whose rows are $\phi_\lambda$s.
Using this eigen-decomposition, 
the PS wavefunction $\tilde{\psi}$  is unitarily transformed to $\tilde{\psi}_{\mathbf{k}\lambda}^{DMFT}$ given by
\begin{eqnarray}
\langle\mathbf{r}|\tilde{\psi}_{\mathbf{k}\lambda}^{DMFT}\rangle & = & \sum_i\langle\mathbf{r}|\tilde{\psi}_{\mathbf{k}i}\rangle
\cdot U_{\mathbf{k}i\lambda}^{DMFT}.
\end{eqnarray}

As a result, the soft charge $\tilde{\rho}$ in Eq.$\:$\ref{eq:eq6} becomes
\begin{eqnarray}
\tilde{\rho}^{DMFT}(\mathbf{r}) & = & \sum_{\lambda}w_{\mathbf{k}\lambda}\langle\tilde{\psi}_{\mathbf{k}\lambda}^{DMFT}|\mathbf{r}\rangle\langle\mathbf{r}|\tilde{\psi}_{\mathbf{k}\lambda}^{DMFT}\rangle.
\end{eqnarray}
while $\bar{\rho}_{mn}$ in Eq.$\:$\ref{eq:eq8} becomes
\begin{equation}
\bar{\rho}_{mn} = \frac{1}{N_{\mathbf{k}}}\sum_{\mathbf{k}\lambda}\langle\tilde{p}_n|\tilde{\psi}_{\mathbf{k}\lambda}^{DMFT}\rangle\cdot w_{\mathbf{k}\lambda}\cdot\langle\tilde{\psi}_{\mathbf{k}\lambda}^{DMFT}|\tilde{p}_m\rangle.
\end{equation}
The final form of the charge density $\rho$ in DFT+DMFT is given by combining Eq.$\:$\ref{eq:eq6}, Eq.$\:$\ref{eq:eq7}, and Eq.$\:$\ref{eq:eq7_2}.
\begin{equation}
\rho^{DMFT}(\mathbf{r}) = \tilde{\rho}^{DMFT}(\mathbf{r}) + \rho^{1,DMFT}(\mathbf{r}) - \tilde{\rho}^{1,DMFT}(\mathbf{r}) 
\end{equation}

\subsection{Full DFT+DMFT self-consistency}

\label{sec:Theory-3}

\begin{figure}[!htbp]
\includegraphics[width=0.9\columnwidth]{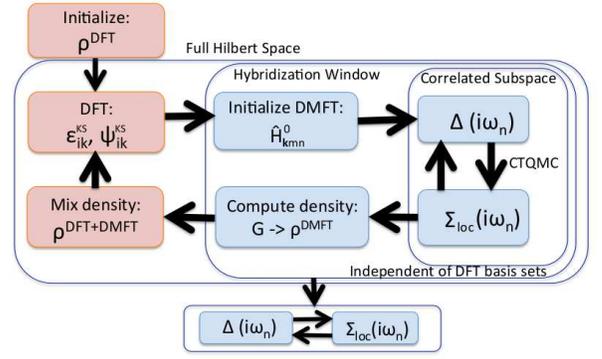}
\caption{(Color online)
Schematic flow diagram of a charge self-consistent DFT+DMFT calculation implemented using the MLWF basis set. The loop is initialized from the solution of the Kohn-Sham equations. The DMFT problem is defined via a set of  Wannier functions and solved as described in Appendix B. The resulting  charge density is computed as explained in Sec.$\:$\ref{sec:Theory-2} and is then used to recompute the single-particle potential and thus the band structure, the charge density  and Wannier functions. Full self-consistency is achieved if both $\rho$ and $G_{loc}$ are converged. For total energies a highly accurate $G_{loc}$ is required, so a post-processing step of approximately 10 DMFT iterations is employed to further refine $G_{loc}$.  
\label{fig:scheme}}
\end{figure}

In this subsection, we present the procedure used to achieve a fully charge
self-consistent solution of the DFT+DMFT equations. A schematic flow diagram
is given in Fig.$\:$\ref{fig:scheme}. 
Full self-consistency is achieved if both $\rho$ and $G_{loc}$ are converged. A highly accurate degreee of convergence is required to obtain reliable results for the total energy. 
The process is normally initialized using
a charge density $\rho$ obtained from the converged non-spin-polarized DFT calculation. This DFT
$\rho$ is a reasonable starting point to obtain a full converged $\rho$ and a
local Green's function $G_{loc}$. Additionally, one must choose a hybridization window, which will encompass the entire $p$-$d$ manifold for the applications
in this paper. 
For the interactions in the correlated subspace, the Slater-Kanamori Hamiltonian is used
with the on-site interaction $U$ and the Hund's coupling $J$ (Eq.$\:$\ref{full_SK}).
Subsequently, the following loop is executed:
\\

\noindent 1.
The DFT potential $V_{DFT}$ is constructed using the input $\rho$ and  GGA,  and the corresponding KS equation is solved
for this given input density $\rho$.
It should be noted that paramagnetic spin symmetry is imposed on the charge $\rho$.
\\

\noindent 2.  MLWF are constructed to represent the KS states in the hybridization window and to construct the correlated subspace (see Sec.$\:$\ref{sec:Theory-1}).
\\

\noindent 3. The DMFT problem is solved to self-consistency using continuous time QMC  to obtain a correlated Green's function $G_{cor}$, a self energy $\Sigma_{cor}$ and a double counting potential $V^{DC}$.  Both $\Sigma_{cor}$ and  $V^{DC}$ are updated at each DMFT step using  linear mixing. Details are given in Appendix B. Obtaining accurate results for the total energy requires a strong convergence of both $\Sigma$ and $V^{DC}$. Convergence is assessed demanding that  $E^{pot}-E^{DC}$ changes by less than 1meV betweeen iterations (see Eq.$\:$\ref{Epot} and Eq.$\:$\ref{eq:DC}). 
\\

\noindent 4. The charge density is constructed from  the Green's function (Eq.$\:$\ref{Gdef}) using the new self energy and the double counting potential (see Sec.$\:$\ref{sec:Theory-2} for definitions). This is then mixed with the previously-computed charge density using  Kerker mixing~\cite{Kerker:81}  in  momentum space, i.e.,
\begin{equation}
\rho(\vec{G})=\rho_{in}(\vec{G})+\alpha\frac{\vec{G}^2}{\vec{G}^2+\gamma^2}(\rho_{out}(\vec{G})-\rho_{in}(\vec{G})).
\label{kerker}
\end{equation}
where $\vec{G}$ is a reciprocal lattice vector while $\alpha$ and $\gamma$ are mixing parameters. 
The new $\rho$ is then returned to step one and this loop is %repeated. % until a desired 
iterated until the change in the charge density at the zone center $\mathbf{k}=0$ satisfies the following criterion. 
\begin{equation}
\frac{1}{N_{\vec{G}}}\sum_{\vec{G}}(\rho_{out}(\vec{G})-\rho_{in}(\vec{G}))^2<10^{-4}~e
\nonumber
\end{equation}

After the DFT+DMFT equations are converged, we rerun the DMFT self-consistent calculation for at least 10 iterations 
using the final charge density. 
In most cases, we were able to converge the total energy to less than 3meV.

\section{Total Energy Calculation}

In this section, we derive the expressions used to evaluate the total energy within our DFT+DMFT formalism in terms of the self-consistent charge density and the local Green's function and self energy obtained as explained in the previous section. Expressions for the double counting energy and potential are  also presented.

\subsection{Formula}
\label{sec:Theory-4}

Our ansatz Eq.~\ref{PhiDFT+DMFT} for the functional implies that the total ground state energy can be written formally as the sum of terms arising from the DFT and DMFT calculations as
\begin{eqnarray}
E^{tot}[\rho,G_{cor}] &=& E^{DFT}[\rho]+E^{KS}[\rho,G_{cor}]
\nonumber \\
&&+E^{pot}[G_{cor}]-E^{DC}[N_d].
\label{eq:energy}
\end{eqnarray}
$E^{DC}$ will be discussed in the next subsection.

$E^{DFT}$ is the energy computed using the conventional DFT formula as
\begin{equation}
E^{DFT}[\rho]=-\frac{1}{2}\sum_i \langle\psi_i|\nabla^2|\psi_i\rangle+\int d\mathbf{r}V_{ext}(\mathbf{r})\rho(\mathbf{r})+E_{Hxc}[\rho].
\label{eq:E_DFT}
\end{equation}

$E^{KS}$ is a correction to the band energy arising the fact that in the hybridization window the density matrix is not equal to the Fermi function. Explicitly,
\begin{equation}
E^{KS} = \frac{1}{N_{\mathbf{k}}}\sum_{\mathbf{k}i}\epsilon^{KS}_{\mathbf{k}i}\cdot(n_{\mathbf{k}ii}- f_{\mathbf{k}i})
\label{eq:E_band}
\end{equation}
where $i$ is a band index, $\epsilon^{KS}_{\mathbf{k}i}$ is the KS eigenvalue, and $n_{\mathbf{k}ii}$  is a diagonal component of the density matrix  computed from $G$ via Eq.$\:$\ref{eq:eq1_1}.

The potential energy  $E^{pot}$ arising from the beyond-DFT interactions in the correlated subspace  is given  by
\begin{equation}
E^{pot}=\frac{1}{2}T\sum_ne^{i\omega_n\cdot 0^-}\mbox{Tr}\left[\hat{\Sigma}_{cor}(i\omega_n)\hat{G}_{cor}(i\omega_n)\right]
\label{Epot}
\end{equation} 
Evaluation of Eq.~\ref{Epot} to the requisite numerical accuracy is facilitated by a careful treatment  of the high frequency limits. Explicitly separating out the leading $\hat{\Sigma}^\infty$  term in the high frequency limit of the self energy for orbital $m$, we define the dynamical self energy by
\begin{equation}
\hat{\Sigma}^{dyn}_m(i\omega_n)= \hat{\Sigma}_m(i\omega_n)-\hat{\Sigma}_m^{\infty}.
\label{Sigmadyn}
\end{equation}
$\hat{\Sigma}_m^{\infty}$ is computed from a Hartree term as derived in Ref.~\onlinecite{Wang:11}.

Substituting Eq.~\ref{Sigmadyn} into Eq.~\ref{Epot} and noting that the frequency summation of the Green's function is just the density matrix $\hat{n}_{cor}$ of the correlated sites gives
\begin{equation}
E^{pot}=\frac{1}{2}\mbox{Tr}\left[\hat{\Sigma}^\infty \hat{n}_{cor}\right]+E^{pot,dyn}
\label{Epot2}
\end{equation}
In evaluating $E^{pot,dyn}$ it is convenient to separate out the very high frequency regime where $\hat{\Sigma}^{dyn}\rightarrow \frac{\Sigma^{dyn,1}}{i\omega_n}$ and $G_{cor}\rightarrow 1/i\omega_n$ by introducing a cutoff frequency $\omega_c$, evaluating frequencies below the cutoff numerically and evaluating the high frequency tail using $\sum_{-\infty}^{\infty} \frac{1}{(2n+1)^2}=\frac{\pi^2}{4}$  to obtain
\begin{eqnarray}
E^{pot,dyn} & \simeq & 
   \frac{1}{2}T\sum_{|\omega_n|<\omega_c}\mbox{Tr}\left[\hat{\Sigma}^{dyn}(i\omega_{n}) \hat{G}(i\omega_{n})\right]\nonumber \\
&  & +\frac{\Sigma^{dyn, 1}}{\pi^{2}T}\left(\sum_{n=0}^{n_{c}}\frac{1}{(2n+1)^{2}}-\frac{\pi^{2}}{8}\right).
 \label{eq:E_pot}
\end{eqnarray}
$\Sigma^{dyn,1}$ is obtained from the expectation value of a combination of
operators; in very simple cases the expectation value
can be computed analytically but in general it must
be measured~\cite{Haule:07,Wang:11}.

For a consistent calculation and the systematic reduction of numerical errors, we compute all static quantities including $\frac{1}{2}\mbox{Tr}[\hat{\Sigma}^{\infty}\cdot \hat{n}_{cor}]$ and $E^{DC}$ (Eq.$\:$\ref{eq:DC}) using the  converged $n_{cor}$ term obtained from the trace of the local Green's function $G_{cor}$. 
We note that the numerical precision is achieved such that 
$n_{cor}$ at each orbital computed from the trace of $G_{cor}$ is converged to
a $n_{cor}$ value sampled from the Monte Carlo method within the numerical error of 0.01.

\subsection{Double counting energy: $U'$ approach}

\label{sec:Theory-5}

The DFT+DMFT ansatz Eq.$\:$\ref{PhiDFT+DMFT} approximates the general functional $\Phi[\rho,G_{cor}]$ as a sum of two terms, one involving $\rho$ only and one involving $G_{cor}$ only.  Such a separation raises the possibility that some interactions will be included in both terms in the sum, and will therefore be counted twice, necessitating the subtraction of an additional `double counting' term to remove the interactions that are counted twice.  In particular the GGA density functional we use to approximate $\Phi_\rho[\rho]$ is a functional of the total charge density, including the charge density in the correlated subspace. Thus some of the interactions contained in $\Phi_\rho$ are also contained in $\Phi_G$ and must be subtracted.   The issue also arises in the DFT+U approximation~\cite{Anisimov:91}.

Determining the double counting energy is not straightforward and, within the approximations adopted above, no exact prescription is known.  However it is essential to address the issue, as the choice of double counting correction affects the energy shift between correlated subspace and the remaining states, which will clearly affect the physics. For example, previous work ~\cite{Wang:12,Park:12,Dang:13,Park:13} has shown that the location of the Mott metal-insulator phase boundary in cuprates~\cite{Wang:12} and early transition metal oxides~\cite{Dang:13} is strongly influenced by the double-counting. Similarly the choice of double counting affects the  bond disproportionation in the rare earth nickelates~\cite{Park:13}.

The double counting correction has been discussed in the literature~\cite{Anisimov:91,Sawatzky:94,Amadon:08,Karolak:10,Wang:12,Dang:13,Haule:13}, mainly in the context of transition metal oxides. Perhaps the most obvious role  that the double counting term plays is in  compensating for the average Hartree shift  of the correlated levels resulting from the interactions in the correlated subspace, as these are present to a large degree within DFT. Stated differently, the splitting between the $d$ and $p$ orbitals within DFT is at least reasonable, and adding an additional Hartree term from the interactions in the correlated subspace would give clearly unphysical results. Additionally, it is reasonable to expect that the double-counting correction should only depend on the total density of the correlated subspace. For these reasons the double counting terms introduced in the literature are typically based on a Hartree approximation to the beyond-DFT interactions in the correlated subspace. A common choice is the fully-localized-limit (FLL) double-counting (defined in Eq.$\:$\ref{eq:DC_FLL}). However, recent studies \cite{Wang:11,Park:12,Dang:13} indicate that in many cases the end result of DFT+DMFT using the FLL double-counting is a relative $p-d$ energy difference in disagreement with experimental photo-emission spectra, while introducing a phenomenological shift to force agreement with a measured level splitting leads to good agreement with a range of other properties~\cite{Dang:13,Park:12}. However,  a phenomenological shift cannot be used in total energy calculations.  In a previous paper~\cite{Park:13} we  introduced a new form of double counting correction, which we refer to as $U^\prime$ double counting, which retains the mathematical form of a standard double counting (and is therefore compatible with total energy calculations) but has an adjustable magnitude. We shall show that this form leads to results in good agreement with experiment, at the expense of the apparent introduction of an additional phenomenological  parameter. We observe that the standard double counting formalisms also involve phenomenological parameters, namely the coefficients multiplying the expectation values of the different density operators. In conventional applications these are set to be equal to the $U$ and $J$ which are used as interactions in the correlated subspace, but this is simply a choice that is made without clear theoretical justification. Therefore, our approach is no more or less phenomenological than the de facto standards in the literature.

Because almost the entire double-counting literature was motivated by the physics of transition metal oxides and the application we present is to transition metal oxides, the rest of our discussion of the double counting correction will refer to these compounds. In this case the correlated orbitals are transition metal $d$-orbitals and the relevant beyond-DFT Hamiltonian can for our purposes be taken to be Eq.~\ref{full_SK}, the  `Slater-Kanamnori' interaction Hamiltonian. We emphasize however that our ideas apply to a wider range of situations, including $f$-electron systems.

We begin with the standard forms of the double counting correction. These are widely employed in the literature.  The philosophy \cite{Anisimov:91} of the double-counting approach is that one should construct a mean-field approximation to the interaction which  depends only on the total occupancy of  the correlated sites (and not, for example on orbital occupancies) since the DFT energy depends only on density. Neglecting the terms which are off-diagonal in the density matrix, the Slater-Kanamori Hamiltonian Eq.~\ref{diag_SK} can be written in terms of the total spin density operator and the Hartree-Fock decoupled orbital dependent terms (ie. $\langle \hat{n}_{\alpha\sigma}\hat{n}_{\beta\sigma'}\rangle=\langle \hat{n}_{\alpha\sigma}\rangle \langle\hat{n}_{\beta\sigma'}\rangle$), leading to
\begin{eqnarray}
E^{cor}_{MF} &=& \frac{U}{2}(N_d^2-\sum_{\alpha\sigma} \langle \hat{n}_{\alpha\sigma}\rangle^2) -\frac{3J}{2}(\sum_{\sigma}N_{d\sigma}^2-\sum_{\alpha\sigma}\langle \hat{n}_{\alpha\sigma}\rangle^2)\nonumber \\ 
&& -J(\sum_{\sigma}N_{d\sigma}N_{d\bar{\sigma}}-\sum_{\alpha\sigma} \langle \hat{n}_{\alpha\sigma}\rangle \langle\hat{n}_{\alpha\bar{\sigma}}\rangle)
\label{eq:E_DC}
\end{eqnarray}

The standard fully localized limit (`FLL') form of the double counting correction assumes that each Fermion at each spin or orbital is either fully occupied or un-occupied (i.e., either zero or one), therefore $\sum_{\alpha\sigma} \langle \hat{n}_{\alpha\sigma}\rangle^2=\sum_{\alpha\sigma} \langle \hat{n}_{\alpha\sigma}\rangle=N_d$.
The final term, $\sum_{\alpha\sigma} \langle \hat{n}_{\alpha\sigma}\rangle \langle\hat{n}_{\alpha\bar{\sigma}}\rangle$ is also approximated as $N_d$ if the paramagnetic constraint is imposed, ie. $\langle\hat{n}_{\alpha\sigma}\rangle=\langle\hat{n}_{\alpha\bar{\sigma}}\rangle$. Finally, the expression of the double counting energy in the paramagnetic state (ie. $N_{d\sigma}=N_{d\bar{\sigma}}=N_d/2$) is given by
\begin{eqnarray}
E^{DC}_{FLL} & = & \frac{U}{2}N_d(N_d-1)-\frac{5J}{4}N_d(N_d-2)
\label{eq:DC_FLL}
\end{eqnarray}
This is identical to the `fully localized limit' scheme~\cite{Anisimov:93,Sawatzky:94,Petukhov:03} modulo the pre-factor of the exchange term. 
The double counting potential $V^{DC}_{\alpha\sigma}$ is given by 
\begin{align}\label{eq:vdc_fll}
V^{DC}_{\alpha\sigma}=\frac{\partial E^{DC}}{\partial n_{\alpha\sigma}}=
U(N_d-\frac{1}{2}) - \frac{5}{2}J (N_d-1)
\end{align}

An alternative form employed in the literature is the ``around mean-field" (AMF) double-counting\cite{Petukhov:03}.
This form is  motivated by  assuming that each orbital has an average occupation of $N_d/10=\langle n\rangle$, resulting in the AMF double-counting energy:
\begin{eqnarray}
E^{DC}_{AMF} & = & \frac{U}{2}N_d(N_d-\langle n\rangle)-\frac{5J}{4}N_d(N_d-2\langle n\rangle)
\label{eq:DC2}
\end{eqnarray}
where $\langle n\rangle=N_d/10$.
This double counting energy gives the AMF double-counting potential:
\begin{align}\label{eq:vdc_amf}
V^{DC}_{\alpha\sigma}=
U(N_d-\langle n\rangle) - \frac{5}{2}J (N_d-2 \langle n\rangle)
\end{align}

Both FLL and AMF double-counting approaches are based on  a double-counting energy which is quadratic in $N_d$ and imply a double-counting potential which is linear in $N_d$. We will also consider an alternative approach  proposed in Ref.~\onlinecite{Haule:10} based on a constant ($N_d$-independent) double counting potential $V^{DC}=\alpha^{dc}$ (in effect a constant level shift) corresponding to 
\begin{eqnarray}
E^{DC}_{shift} & = & \alpha^{dc} N_d.
\label{eq:DC4}
\end{eqnarray}
This is not an interaction energy, because the energy is linear in $N_d$ as opposed to quadratic. 
All three forms will be considered in this study.

The AMF and FLL  double-counting corrections have difficulties when compared in detail to experiment, in particular  producing a $V^{DC}$ that leads to a $d$-$p$ level separation which is in disagreement with experiment\cite{Wang:12,Dang:13}.   It seems desirable to design a double-counting energy which has the form of an interaction energy but which permits modifications of $V^{DC}$. In previous work ~\cite{Park:13} we proposed a modification that fulfils these criteria,  changing the coefficient  $U$ value in the  double counting formula to a new coefficient  $U'$ while otherwise leaving the form unchanged. We refer to this as the $U^\prime$ double-counting approach, and it may be applied equally well to both the FLL and AMF formulas. In the case of the FLL double-counting, we have explicitly
\begin{eqnarray}
E^{DC}_{FLL} & = & \frac{U'}{2}N_d(N_d-1)-\frac{5J}{4}N_d(N_d-2)
\label{eq:DC}
\end{eqnarray}
More generally, one could consider an arbitrary quadratic double-counting correction, modifying $J$ also or introducing additional terms, but because the $U^\prime$ approach has proven to be satisfactory~\cite{Park:13} we have not explored these changes.  It should be noted that the $J$-term does change the ratio of the linear and quadratic terms as compared to the $U^\prime$ term, but this is not critical to achieving the proper physics.

In a similar way, the modified AMF formula is given by
\begin{eqnarray}
E^{DC}_{AMF} & = & \frac{U'}{2}N_d(N_d-\langle n\rangle)-\frac{5J}{4}N_d(N_d-2\langle n\rangle).
\label{eq:DC_AMF}
\end{eqnarray}
We will see that a single $U^\prime \neq U$  produces good results for the phase diagram and spectra across an entire family of nickelate compounds addressed in this paper.

\section{Application to rare earth nickelates}
\label{sec:Results}

\subsection{Overview}
\label{sec:Results-1}

We use the DFT+DMFT total energy implementation developed here
to calculate the metal-insulator and structural phase diagrams of the rare earth nickelate
family of materials in the plane of rare earth ion (“tolerance factor”) and pressure. 
We present results for bond lengths, electron spectra and other properties as well.

The rare earth nickelates are a family of materials with chemical formula $R$NiO$_3$ with $R$ being a member of the rare earth series La, Nd, Pr, Sm, Eu, and Lu. The important electronic degrees of freedom reside in the Ni $d$-levels. Formal valence arguments indicate that the Ni is dominantly in the $d^7$ configuration, with filled O $2p$ and Ni $t_{2g}$ shells and one electron in the $e_g$-symmetry ($d_{x^2-y^y}$ and $d_{3z^2-r^2}$) states. However the $d$-levels are very strongly hybridized to the O $2p$ states so the actual configuration is much closer to $d^8$ with a hole on the oxygen bands. At high temperatures all of the members of the series are metallic and except for LaNiO$_3$ all crystallize in a $Pbnm$ structure which is derived from the standard cubic perovskite structure by octahedral rotations (LaNiO$_3$ forms in a $R\bar{3}c$ structure also derived by rotations from the cubic perovskite). All of the members of the series except for $R$=La undergo a metal to insulator transition as temperature is decreased at ambient pressure, but at low  temperature the metallic phase may be restored by application of sufficient pressure \cite{Medarde:1997,Zhou:04,Amboage:05,Amboage:04,Ramos:12}. The metal to insulator transition, driven by beyond band theory electron correlations that produce  an unusual site-selective Mott insulating phase \cite{Park:12},  is intimately coupled with a transition to a $P2_1/n$ structure characterized by a two-sublattice bond disproportionation in which one of the  Ni has a short   mean Ni-O bond length while the other has long mean Ni-O bond length\cite{Medarde:97}. The amplitude of the bond-length disproportionation and the critical pressure required to restore the metallic phase depend on the choice of rare earth ion $R$. Thus the behavior of this class of materials is determined by a sensitive interplay of structural and correlated electron physics and  presents a significant test for a theory of correlated electron materials.

\subsection{Formalism and computational aspects}
\label{sec:Results-2}

The DFT portion of the formalism is solved using the PAW formalism~\cite{Blochl:1994,Kresse19991758} as implemented in Vienna Ab-initio Simulation Package (VASP)~\cite{PhysRevB.49.14251,PhysRevB.47.558,Kresse199615,Kresse199611169,Kresse19991758}. A $k$-point mesh of $6\times6\times6$ (for the $Pbnm$ and $P2_1/n$ structures) or $8\times8\times8$ (for the LaNiO$_3$ $R\bar{3}c$ structure) is used with an energy cutoff of 600eV.  
We used Kerker mixing parameters of $\gamma$=1.0 and $\alpha$=0.1 (see Eq.$\:$\ref{kerker}), which resulted in  slow but stable convergence.  
The hybridization window is taken to span the manifold of Ni-$d$ and O-$p$ states which has a range of roughly 11eV, and the correlated subspace is constructed as outlined in Section \ref{sec:Theory-1}.  The interactions pertaining to these orbitals are given by the rotationally invariant Slater-Kanamori Hamiltonian  including the on-site intra-orbital Coulomb interaction $U$ and the Hund's coupling $J$:
\begin{eqnarray}
\label{full_SK}
\hat{H}_{cor} &=& \hat{H}_{D} + \hat{H}_{OD} \\
\label{diag_SK}
\hat{H}_{D} &=& U\sum_{i,\alpha}\hat{n}_{i\alpha\uparrow}\hat{n}_{i\alpha\downarrow}
+(U-2J)\sum_{i,\alpha\neq\beta}\hat{n}_{i\alpha\uparrow}\hat{n}_{i\beta\downarrow} \nonumber \\ 
& & +(U-3J)\sum_{i,\alpha>\beta,\sigma}\hat{n}_{i\alpha\sigma}\hat{n}_{i\beta\sigma} \nonumber \\
&=& \frac{1}{2}U\sum_{i}(\hat{N}_{d,i}^2-\hat{N}_{d,i}) %\nonumber \\
-\frac{3}{2}J\sum_{i,\sigma}(\hat{N}_{d,i\sigma}^2-\hat{N}_{d,i\sigma}) \nonumber \\
& & -J\sum_{i,\sigma}(\hat{N}_{d,i\sigma}\hat{N}_{d,i-\sigma}-\sum_{\alpha}\hat{n}_{i\alpha\sigma}\hat{n}_{i\alpha-\sigma}) \\ 
\hat{H}_{OD} &=& J\sum_{i,\alpha\neq\beta}\left( \hat{\psi}^{\dagger}_{i\alpha\uparrow}\hat{\psi}_{i\beta\uparrow}
\hat{\psi}^{\dagger}_{i\beta\downarrow}\hat{\psi}_{i\alpha\downarrow} %\nonumber \\
 +\hat{\psi}^{\dagger}_{i\alpha\uparrow}\hat{\psi}_{i\beta\uparrow}
\hat{\psi}^{\dagger}_{i\alpha\downarrow}\hat{\psi}_{i\beta\downarrow} \right) \nonumber  \\
\label{eq:Eq_cor}
\end{eqnarray}
where $i$ is the Ni atom index, $\alpha$ is the $d$ orbital index, and $\sigma$ is the spin.
$\hat{N}_{d,i}(=\sum_{\alpha,\sigma}\hat{n}_{i\alpha\sigma})$ is the total $d$-occupancy operator acting on the Ni atom $i$.

Unless otherwise specified the computations in this section are performed for $U=5eV$ and $J=1eV$ and the double counting correction is the FLL-$U^\prime$ form of Eq.~\ref{eq:DC} with $U^\prime=4.8eV$.

Since the $t_{2g}$ orbitals are almost filled, they are approximated using the Hartree-Fock approximation while the self energy of $e_g$ orbitals is obtained using the single-site dynamical mean field approximation\cite{Georges:96} with the numerically exact `continuous-time QMC method'~\cite{Werner:06,Werner:2006,Haule:07,Gull_review:11}. With this technique, temperatures as low as $0.01eV$ are  accessible, low enough that the energies we calculate are representative of the ground state energy.  Details of the DMFT procedure are given in Section \ref{sec:Theory} and Appendix B.

\begin{figure}[t]
\includegraphics[width=0.82\columnwidth]{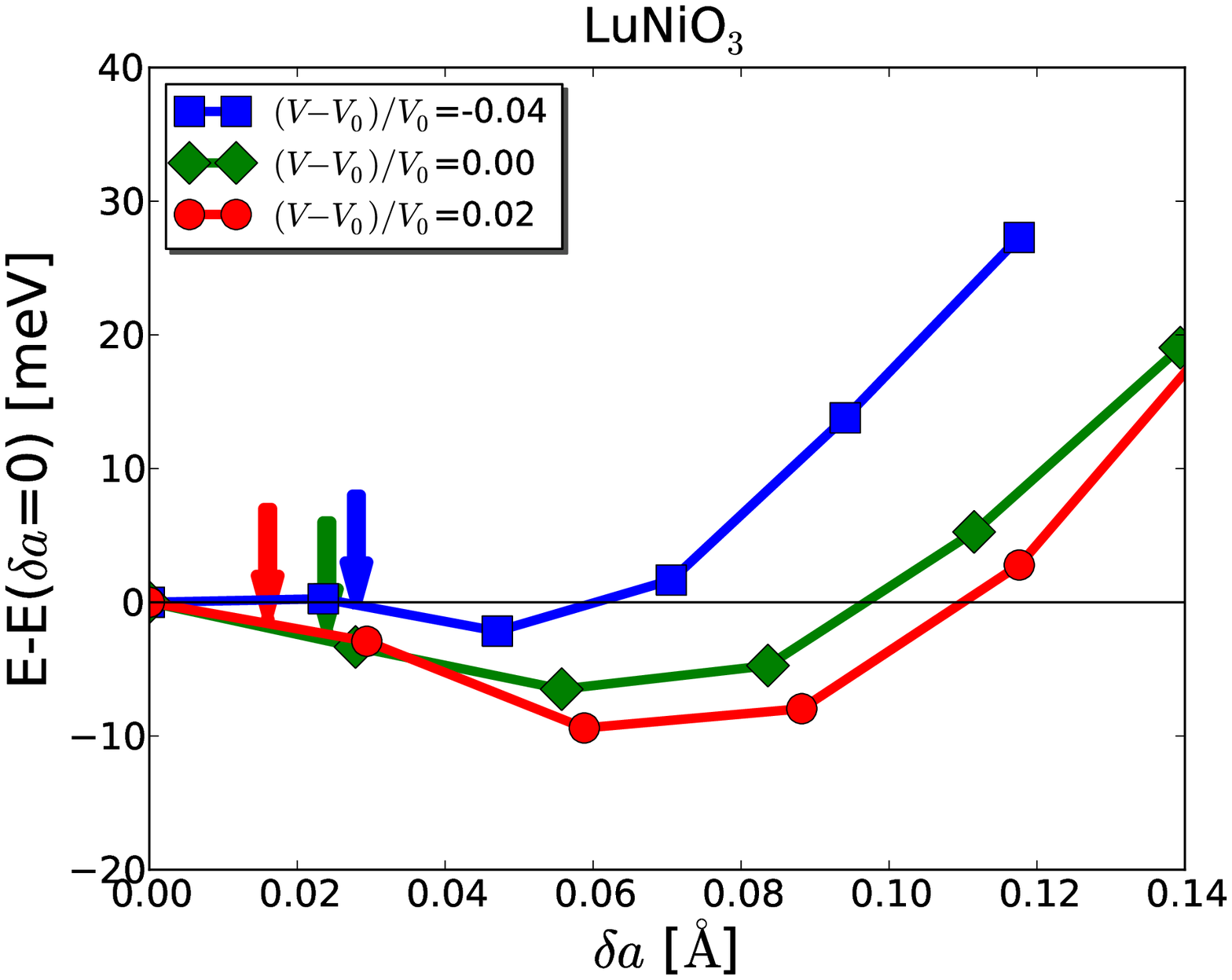}
\includegraphics[width=0.82\columnwidth]{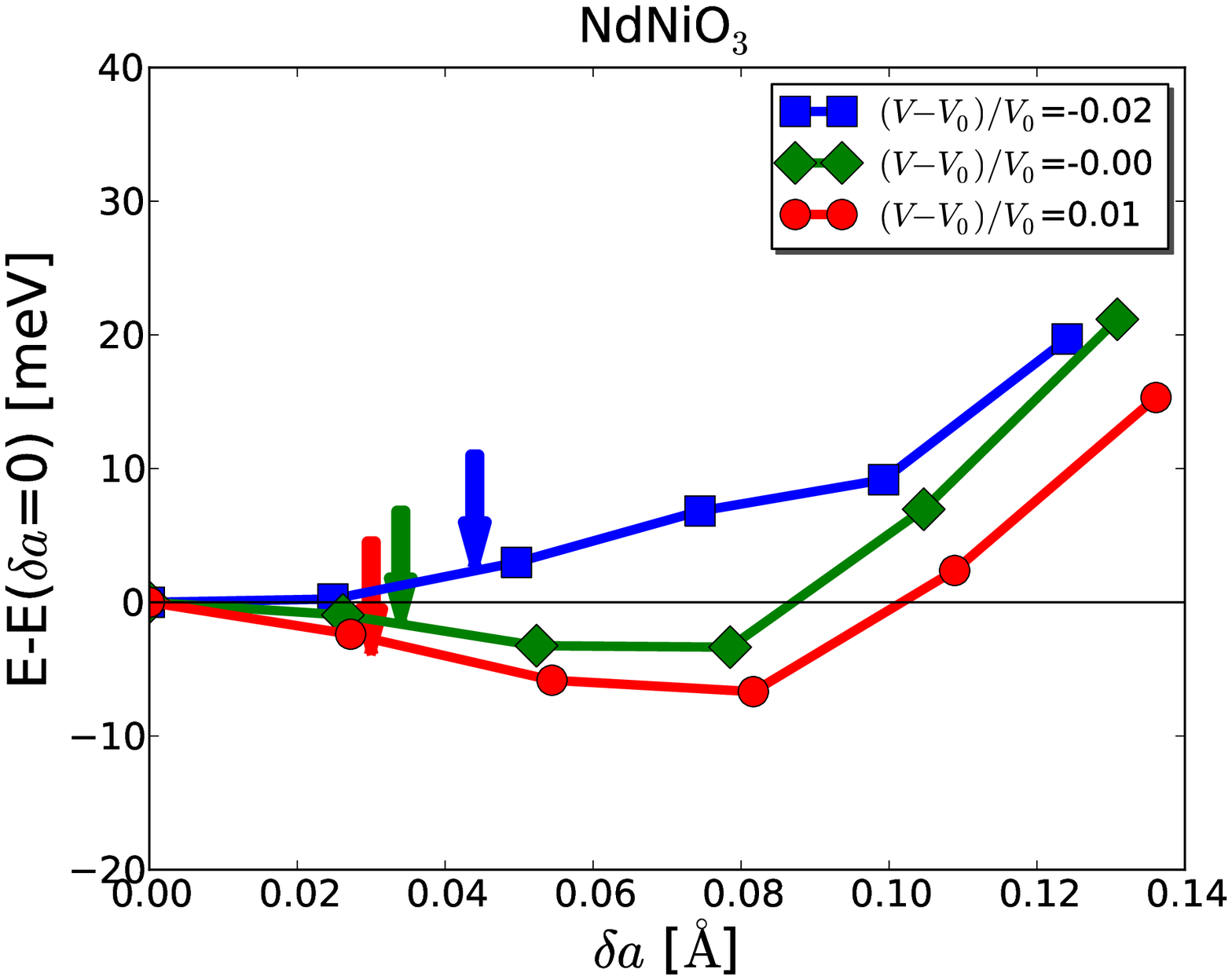}
\includegraphics[width=0.82\columnwidth]{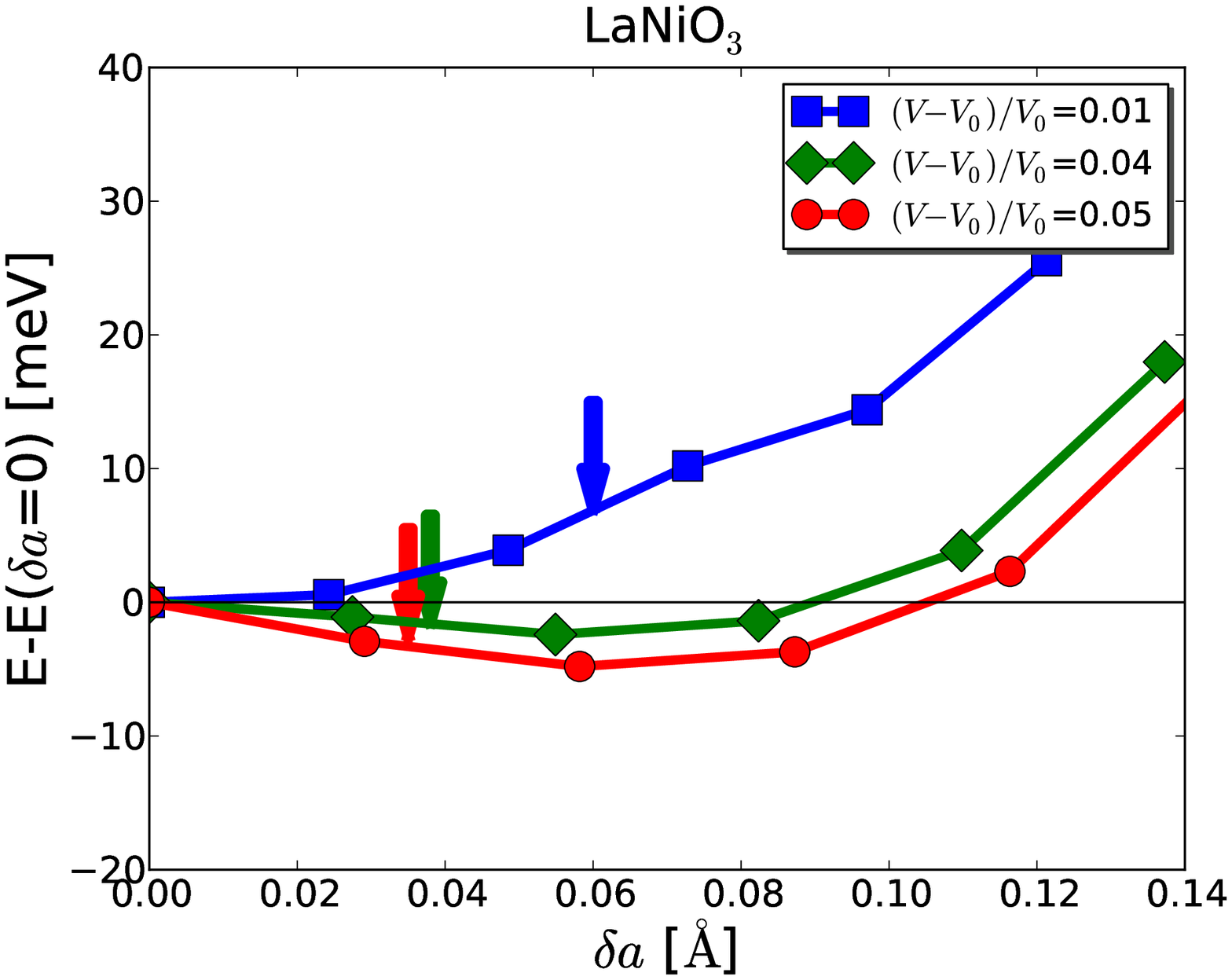}
\caption{(Color online) 
Total energy as a function of bond-length difference $\delta a$ for LuNiO$_3$ (top), NdNiO$_3$ (middle) and LaNiO$_3$ (bottom) obtained from DFT+DMFT calculations performed at different values of the unit cell volume $V$ measured relative to the calculated zero pressure volume $V_0$. Arrows indicate the $\delta a$  at which the electronic phase changes from metal to  insulator. \label{fig:E_DMFT}}
\end{figure}

In order to determine the theoretical structure one needs to minimize the energy over the space of possible structures. We have not yet implemented the computation of forces and stresses within our formalism  and a direct minimization of the energy via exploration of the entire space of structures would greatly exceed our computational resources. Therefore, we approximately minimize the energy via the construction of a two dimensional phase space parametrized by unit cell volume and Ni-O bond length disproportionation. To define the phase space we  use  the VASP implementation of  DFT to determine the internal coordinates and cell shape that minimize the energy consistent with the known symmetries of the high temperature ($Pbnm$) phase at each volume. Similarly we use the VASP implementation of DFT+U, which uses projectors to construct the correlated subspace, to find the internal coordinates and cell shape consistent with the $P2_1/n$ symmetry of the low temperature phase at a given volume. It should be noted that these VASP DFT+U calculations use a spin independent exchange-correlation functional and double-counting formula to compute the total energy, which is analogous to our DFT+DMFT formalism (this can be achieved in VASP by setting the LDAUTYPE tag to be 4).
At each volume, a one-dimensional path is determined by interpolating from the $Pbnm$ structure to the distorted $P2_1/n$ structure and is parametrized by the mean Ni-O bond length difference $\delta a$ between the two inequivalent sub-lattices. At each  volume, the $\delta a$ value is obtained by minimizing the total energy along this one dimensional path (see Fig.$\:$\ref{fig:E_DMFT}). 
The same procedure was adopted for LaNiO$_3$ to determine the two dimensional phase space except that LaNiO$_3$ is based on the $R\bar{3}c$ structure. 
The structural phase boundary is defined by the volume at which the minimum of the energy curve moves away from $\delta a=0$ (in practice,  $\delta a > 0.01 \AA$).

\begin{figure}[b]
\includegraphics[scale=0.45]{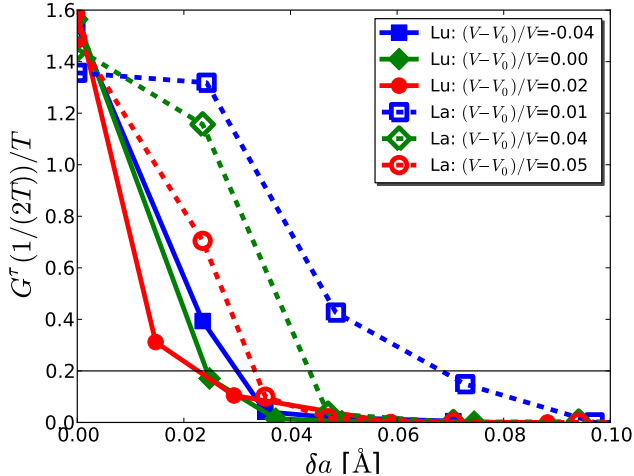}
\caption{(Color online) 
The many-body density of states at the Fermi energy (averaged per Ni atom) for LuNiO$_3$ (solid lines, filled dots) and LaNiO$_3$ (dashed lines, empty dots)  as a function of $\delta a$ computed using DFT+DMFT at volumes indicated.  The horizontal line at y=0.2 indicates the criteria for the metal-insulator transition.
 \label{fig:A0}}
\end{figure}

We define the electronic phase as insulator or  metal according to whether the  electron spectral function (imaginary part of the real-frequency local Green's function) has a gap at the Fermi level  or not. For computational convenience and to avoid the errors associated with analytical continuation we employ the relation \cite{Werner098site} (the rightmost approximate equality becomes exact as temperature $T\rightarrow 0$) 
\begin{equation}
\frac{G(\tau=1/(2T))}{T} = \int \frac{d\omega}{\pi T}\frac{A(\omega)}{2\cosh{\frac{\omega}{2T}}}\simeq A(\omega=0)
\label{findA}
\end{equation}
between the Green's function measured in imaginary time by the continuous time QMC procedure and the Fermi-level spectral function of interest.  
In a Fermi liquid at $T=0$ within the single-site DMFT approximation,
$A(\omega=0)$ is of the order of the bare Fermi level density of states 
(the Hartree-like shift of the $d$-$p$ energy level difference arising
from $Re\Sigma(\omega=0)$ will alter the band structure,
therefore even in the Fermi liquid regime $A(\omega=0)$ is not equal to the bare Fermi level density of states).
Numerically, we define a material as a metal if $A(\omega=0)$ computed by Eq.~\ref{findA} is greater than $0.2$ and as insulator if $A(\omega=0)<0.2$. While the criterion is not completely precise, it is fully adequate for our purposes. Examples of the dependence of $G(\tau=1/(2T))/T\approx A(\omega=0)$ in LuNiO$_3$ and LaNiO$_3$ are given in Fig.$\:$\ref{fig:A0}.

\subsection{Phase diagram: Pressure vs rare-earth ion series}
\label{sec:Results-3}

\begin{figure}[b]
\includegraphics[scale=0.45]{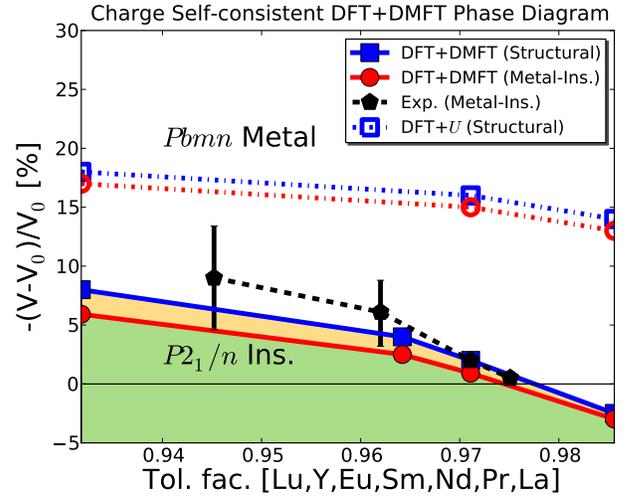}
\caption{(Color online)
Metal-insulator (circle dots) and structural (square dots) phase diagram computed using charge self-consistent DFT+DMFT (solid symbols and solid lines) as a function of volume (y-axis) and the series of rare-earth ions (x-axis). The tolerance factor is defined as $d_{R-O}/d_{Ni-O}\sqrt{2}$ where $d_{R-O}$ and $d_{Ni-O}$ are $R$-O and Ni-O distances~\cite{Medarde:97}. Experimental data (pentagons and dashed lines, black on-line) are obtained for (Y,Eu,Nd,Pr)NiO$_3$ using the data Ref.$\:$\onlinecite{Cheng:10} as explained in the text.  DFT+U results (empty symbols and dashed lines) using the same correlated orbital are also compared to DFT+DMFT results. $V_0$ is determined as equilibrium volume at the calculated zero pressure for each material using each theoretical method. The parameters for the DFT+DMFT calculations are $T$=116K, $U$=5eV and $J$=1eV. The double counting energy is determined using Eq.$\:$\ref{eq:DC} with $U'$=4.8eV. The DFT+U calculations are performed using $U$=5eV, $U'$=5eV and $J$=1eV.
\label{fig:phased}}
\end{figure}

Fig.$\:$\ref{fig:phased} shows the  metal-insulator (circle dots, filled symbols) and structural (square dots, filled symbols) phase transitions computed from charge self-consistent DFT+DMFT (solid lines, filled symbols) as described above, in addition to DFT+U results (dashed lines, open symbols). The experimental critical phase boundaries (pentagons and black dashed lines) for (Y,Eu,Nd,Pr)NiO$_3$ obtained from Ref.$\:$\onlinecite{Cheng:10} are also shown. These are determined   using extrapolation of high temperature experimental data to low temperature as  explained in Ref.$\:$\onlinecite{Park:13}.  The theoretical DFT+DMFT metal-insulator transition phase diagram (solid lines and circle dots) in Fig.$\:$\ref{fig:phased} is consistent with experimental data (black dashed lines and pentagons) in agreement with our previous non-charge self-consistent calculation\cite{Park:13}. The two key points of comparison with experiment are:

\begin{enumerate}
\item all nickelates calculated at  zero pressure are insulating and bond-length disproportionated
except  LaNiO$_3$ which remains metallic and un-disproportionated. 

\item the critical pressure line at which the insulator-to-metal transition occurs is  quantitatively in  good agreement with experiment. Stated differently, the critical volume becomes larger as the rare-earth ion size increases from Lu to La such that LuNiO$_3$ requires nearly 6\% contraction of volume to induce the insulator-to-metal transition while 3\% volume expansion of LaNiO$_3$ would exhibit a metal-to-insulator transition (including bond-disproportionation).
\end{enumerate}

LaNiO$_3$ is the only nickelate with a rhombohedral structure and at the zero pressure exhibits a metallic ground state without any bond disproportionation, consistent with the experimental observation.

Comparison to the non-charge self-consistent results presented in Ref.~\onlinecite{Park:13} shows that  charge self-consistency systematically shifts the phase boundary towards large volume and a smaller rare-earth ion size, decreasing the regime of insulating behavior. The physical origin is due to the slightly reduced $d$-$p$ gap and therefore the reduction of electronic correlations in charge self-consistent calculations compared to the non-charge self-consistent ones.

We have also computed the phase boundary using the DFT+U approximation (dashed lines, open symbols).  These computations use the same correlated orbital (MLWF) and same spin-independent exchange-correlation function as was used in our  DFT+DMFT calculations, and minimize the total energy in the same two-dimensional phase space of volume and $\delta a$. The only difference between the two calculations is that the DFT+U calculation solves the many-body problem with a Hartree approximation.  Ensuring that the two calculations are built on the same foundation is important for a clear comparison, as may be demonstrated by examination of the DFT+U phase diagram previously reported in Ref.$\:$\onlinecite{Park:13}. The previous computation used the conventional VASP DFT+U implementation, based on a spin-dependent exchange-correlation functional, which is a different approximation leading to significant differences in the results. Additionally, in the previous computation the correlated subspace was constructed using projectors rather than Wannier functions. Examination of the effects of choice of the exchange-correlation potential and methodology for constructing the correlated subspace is beyond the scope of this paper and will be considered elsewhere \cite{Park:14}. What is important for this paper is that the DFT+U lines in Fig. ~\ref{fig:phased} clearly demonstrate the poor quality of the Hartree approximation, which  strongly overestimates the tendency to insulating behavior and  charge disproportionation, predicting for example that  LaNiO$_3$ at the zero pressure is insulating and bond-disproportionated in clear disagreement with the experiment. Another deficiency of the DFT+U approach is that the critical volume is predicted to change much more slowly with rare earth ion than is observed or calculated with DFT+DMFT.

\subsection{Bond-length difference $\delta$a vs. pressure}
\label{sec:Results-4}

\begin{figure}[!htbp]
\includegraphics[scale=0.45]{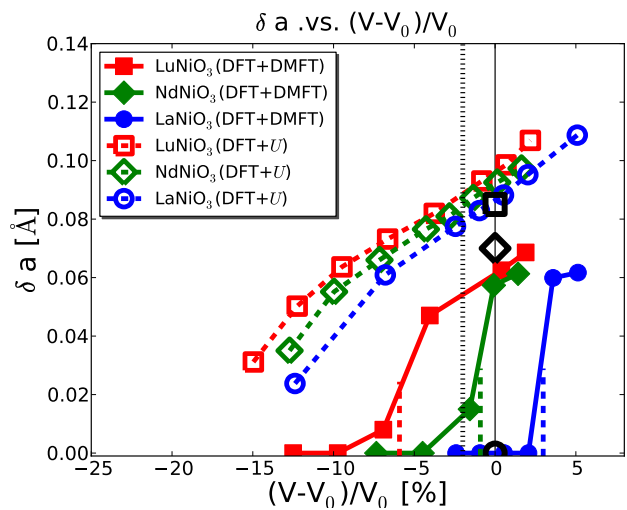}
\caption{(Color online) 
The average Ni-O bond-length difference $\delta a$ of the two inequivalent octahedra of the $P2_1/n$ structure as a function of the normalized difference of the volume $V$ from the zero pressure volume $V_0$ for LuNiO$_3$ (red square), NdNiO$_3$ (green diamond),  and LaNiO$_3$ (blue circle) calculated using DFT+DMFT  (solid symbols, solid lines), DFT+$U$ results (open symbols, dashed lines) and experimental data (black open symbols)~\cite{Alonso:01,Garcia:92,Garcia:09} at ambient pressure. The vertical black dotted line shows the reduced volume for NdNiO$_3$ at which the experimental metal-insulator transition occurs. The theoretical critical volumes at which the metal-insulator transition occurs are depicted as vertical dashed lines connecting to the different $\delta a$ curves.
 \label{fig:bondvsvol}}
\end{figure}

Fig.$\:$\ref{fig:bondvsvol} displays the ground state $\delta a$ calculated at different volumes  using DFT+DMFT and DFT+U (dashed lines, open symbols)   for LuNiO$_3$, NdNiO$_3$,  and LaNiO$_3$ (blue circle dots). Experimental results at ambient pressure are  shown as open symbols and are in reasonable agreement with the DFT+DMFT predictions for $\delta a$. For example, the calculated zero pressure $\delta a$ for LuNiO$_3$ is $\sim 0.065\AA$, slightly less than the experimental value $\sim0.085\AA$ (black open  square) while the calculated value for NdNiO$_3$ is $\sim 0.06\AA$  only slightly smaller than the experimental value $\sim0.07\AA$ (black diamond dot).
As pressure is applied (volume is reduced), $\delta a$ decreases and then sharply drops  at the insulator to metal transition (labelled by the vertical dashed lines). 
The calculated critical volume at which the metal insulator transition occurs  in NdNiO$_3$ (green vertical dashed line) is slightly larger than the experimental volume (black vertical dashed line). 
In LaNiO$_3$ (rhombohedral structure) the DFT+DMFT calculation predicts undistorted ($\delta a=0$) metallic behavior in agreement with experiment.  In contrast, DFT+U qualitatively fails to reproduce the properties of LaNiO$_3$ at the calculated zero pressure,  predicting instead a large $\delta a$ $\sim$0.09$\AA$ and an insulating ground state. DFT+U overestimates the $\delta a$ values at the calculated zero pressure  for all other nickelates as well, consistent with the error in critical volume reported in  Fig.$\:$\ref{fig:phased}.

The physical origin of this behavior can be understood. As pressure increases (smaller volumes, square dots), the critical $\delta a$ required to drive an insulating state increases for both LuNiO$_3$ and LaNiO$_3$, essentially because at smaller cell volume the hybridization (kinetic energy) increases so the electrons are relatively less correlated. LaNiO$_3$ has larger critical $\delta a$ values at the same pressure than LuNiO$_3$  because a structural difference (more nearly straight O-Ni-O bond) means that the bandwidth of the La compound is greater than that of the Lu compound.

\subsection{Double counting}
\label{sec:Result_dc-1}

In all preceding calculations we presented results generated using $U=5eV$ and the FLL-$U^\prime$ double counting, Eq.$\:$\ref{eq:DC},  with $U'=4.8eV<U$.  This choice of double counting differs from the standard FLL double counting procedure which in our notation corresponds to  $U'=U$. In the following subsections we examine the consequences of choosing different values of $U^\prime$ and   provide a more detailed discussion of  how we arrived at the value of $U'=4.8eV$, showing in particular that it produces spectra in better agreement with experiment.  We also present results obtained by other double counting procedures.

In the following subsections we use non-charge self consistent calculations. The reason is that in transition metal oxides the double counting correction acts to shift the energy of the $d$-levels relative to that of the $p$-levels. The charge self-consistency procedure also has the effect of shifting the $p$-$d$ energy splitting and interacts in a non-linear way with the changes induced by the double counting correction. Thus to isolate the effect of the double counting correction, in the following subsections only  we do not include charge self consistency.

\subsection{Varying  $U'$}
\label{sec:Result_dc-2}

In this subsection, we compare disproportionation amplitudes $\delta a$  obtained using different double countings. We also present some results for the electron spectral function.  In particular, we  demonstrate that the location of the phase boundary depends on the choice of double counting correction and that the conventional choice $U'=U$ gives a qualitatively wrong result.

\begin{figure}[!htbp]
\includegraphics[scale=0.45]{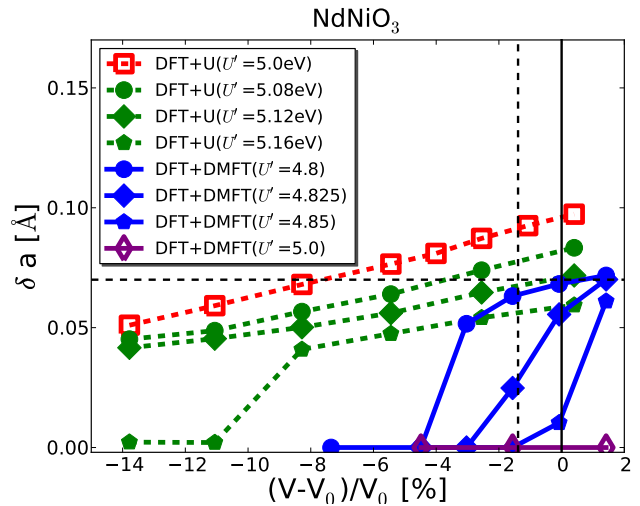}
\caption{(Color online)
The bond-length difference $\delta a$ for NdNiO$_3$ as a function of unit cell volume $V$ measured relative to calculated equilibrium volume $V_0$ computed  using DFT+DMFT (solid lines) and DFT+$U$ (dashed lines) with $U^\prime$-FLL double counting and $U^\prime$ values indicated. The horizontal dashed line indicates the experimental $\delta a$ value for NdNiO$_3$. \label{fig:bondUp}}
\end{figure}

Fig.$\:$\ref{fig:bondUp} displays $\delta a$ values for NdNiO$_3$  as a function of unit cell volume computed using the $U^\prime-FLL$ double counting formula Eq.~\ref{eq:DC} with different choices  of $U^\prime$. We see that results depend on the value of $U^\prime$, with the bond-length disproportionation systematically decreasing as $U^\prime$ is increased in both the DFT+DMFT and the DFT+$U$ calculations. The dependence of result on $U^\prime$ demonstrates the importance of employing a correct double counting term.

The physics of the $U^\prime$ dependence is that in transition metal oxides the degree of correlation is controlled to a large degree by the energy difference between the oxygen $p$ and transition metal $d$ levels. The higher the $d$-levels are above the $p$-levels, the more strongly correlated the material is.  Because the double counting correction enters with a negative sign, increasing $U^\prime$ acts to shift the $d$-levels down relative to the $p$-levels, thereby decreasing the correlation strength. 
The nominal $p-d$ splitting can equivalently be characterized by the number of electrons in the correlated subspace $N_d$.
We found that the metal-insulator phase diagram in the plane of $U$-$N_d$ takes a simple and general form, with the system becoming less correlated as $N_d$ increases and displaying a threshold behavior whereby an insulating state cannot be achieved beyond a certain value of $N_d$ for any practical $U$\cite{Wang:12,Dang:14a,Dang:14b}. To a large degree differences between different methodologies (charge self consistent or not, different forms of double counting correction) disappear when the results are expressed in terms of $N_d$; in other words the main reason for differences between different methodologies is the difference in the relation between $N_d$ and the bare parameters of the theory. It should be noted that the absolute value of $N_d$ depends upon the details of the definition of the correlated subspace, but the relative differences from the DFT value provide a useful representation of the physics. Using our Wannier construction of the correlated subspace, DFT calculations for the rare earth nickelates lead to $N_d\sim 8.2$. Our non-charge self-consistent DFT+DMFT calculations with FLL double counting at $U^\prime=U=5eV$ yield $N_d\approx 8.07$ and predicts a  $Pbnm$ structure and metallic ground state at zero pressure for all members of the series, exhibiting a similar qualitative failure to standard DFT. %Therefore, it is necessary to more generally understand how the double-counting is affecting the results.

Fig.$\:$\ref{fig:bondUp}  also displays results obtained with the DFT+U method, using the same double counting. As is to be expected from the results already presented, DFT+$U$ with the standard FLL $U^\prime=U$  double counting greatly overestimates the calculated zero pressure $\delta a$. Decreasing $U^\prime$ relative to $U$ only worsens the disagreement with experiment. %One could empirically consider whether DFT+U could be enhanced by reducing the effective correlations via increasing $U^\prime$ and hence increasing $N_d$. For example, 
One could consider increasing $U^\prime$ relative to $U$. This  of course reduces the $\delta a$ value while increasing $N_d$. For NdNiO$_3$ the calculated $\delta a$ becomes similar to experiment at $U'$=5.12eV ($N_d$=8.24). However, the critical pressure for the structural transition is still grossly overestimated even at $N_d$=8.24 $(-(V-V_0)/V_0>14\%)$ and other aspects of the physics such as the $p-d$ energy splitting are wrongly predicted as compared to experiment. We will  demonstrate below that in the interacting theory $N_d$ should be reduced relative to the DFT value, while increasing $U^\prime$ relative to $U$ has the opposite effect. Thus we believe that increasing $U^\prime$ in the DFT+U formalism amounts to correcting the errors of the Hartree approximation by introducing a new error.

We now turn to a different observable, the electron spectral function, which has features revealing the energy positions of the oxygen $p$ and transition metal $d$ states, and can be measured in photo-emission and resonant X-ray scattering experiments.

\begin{figure}[!htbp]
\includegraphics[scale=0.4]{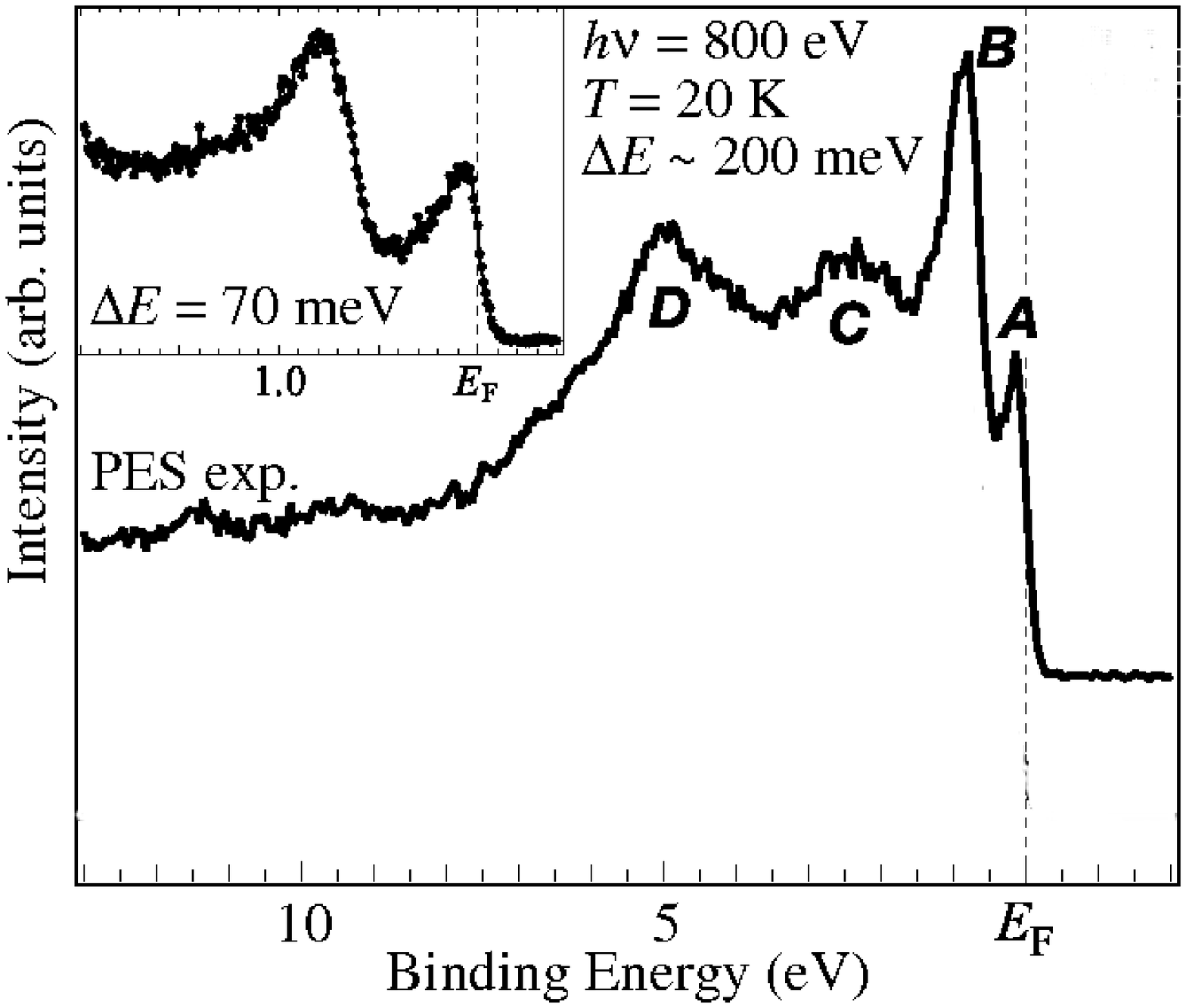}
\includegraphics[scale=0.45]{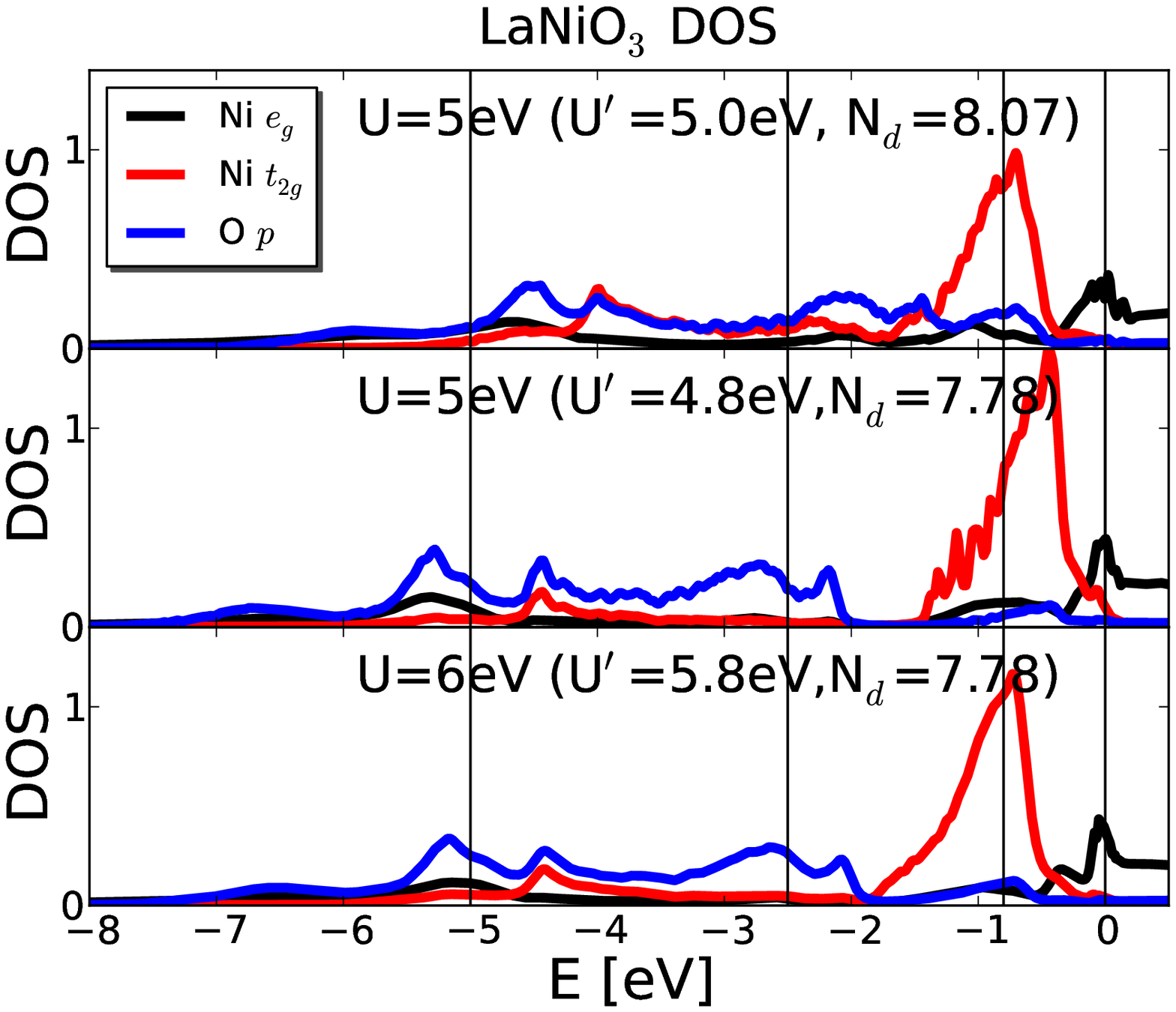}
\caption{(Color online)
Upper panel: The experimental  photo-emission spectra measured in a LaNiO$_3$ thin film~\cite{Horiba:07} with $Ni$ $e_g$ (A) and $t_{2g}$ (B) $d$ states and oxygen $p$ (C,D) features identified. Lower panel: DFT+DMFT spectral functions in LaNiO$_3$  computed using the experimental  $R\bar{3}c$  structure using different values of $U'$ and $U$. Note the difference in energy scale relative to the top panel. The black horizontal lines show the energies of the peaks  indicated in the upper panel.
\label{fig:DOS}}
\end{figure}

The top panel of Fig.$\:$~\ref{fig:DOS} shows the experimental photo-emission spectra~\cite{Horiba:07} of thin film LaNiO$_3$. The peaks A and B correspond to the Ni $e_g$ and $t_{2g}$ states, respectively, and the peaks C and D represents O $p$ states. The bottom panel of Fig.$\:$~\ref{fig:DOS} displays orbitally resolved DFT+DMFT spectral functions  calculated using $U=5eV$ and $U^\prime=5eV$ and $4.8eV$.  We see that the conventional double counting $U^\prime=5eV$ places the oxygen peaks at noticeably higher energies than is compatible with the data.  This error in the oxygen energy corresponds to a larger $N_d$ and effectively weaker correlations, explaining the lack of disproportionation predicted by this double counting. By contrast the $U^\prime=4.8eV$ double counting places the oxygen bands at approximately the correct energy. Although the correspondence between calculated and experimental spectra is not perfect, and could be improved by further fine-tuning, it is clear that the shift induced by reducing  $U^\prime$ relative to $U$ is physically reasonable and produces both basically correct spectra and a reasonable structural phase diagram. It should be noted that the $t_{2g}$ has shifted slightly above the experimental peak when going from $U^\prime=5.0eV$ to $U^\prime=4.8eV$. Better agreement of the $t_{2g}$ state can be regained  without compromising the O $p$ peaks by increasing $U$, using $U=6eV$ and $U'=5.8eV$ (see Fig.$\:$~\ref{fig:DOS} bottom panel). However the $t_{2g}$ states are filled and their exact placement is not relevant to the physics of the site-selective Mott transition. The uncertainties induced by the other approximations inherent in the DFT+DMFT procedure suggest that further fine-tuning to bring the oxygen spectra into even better alignment with the data is not warranted at this time. 

%Our smaller value of $U'$=4.8eV also reduces the $d$ occupancy $N_d$ from 8.07 ($U'$=5eV) to 7.78.

\subsection{Different double counting formulae}
\label{sec:Result_dc-3}

\begin{figure}[!htbp]
\includegraphics[scale=0.45]{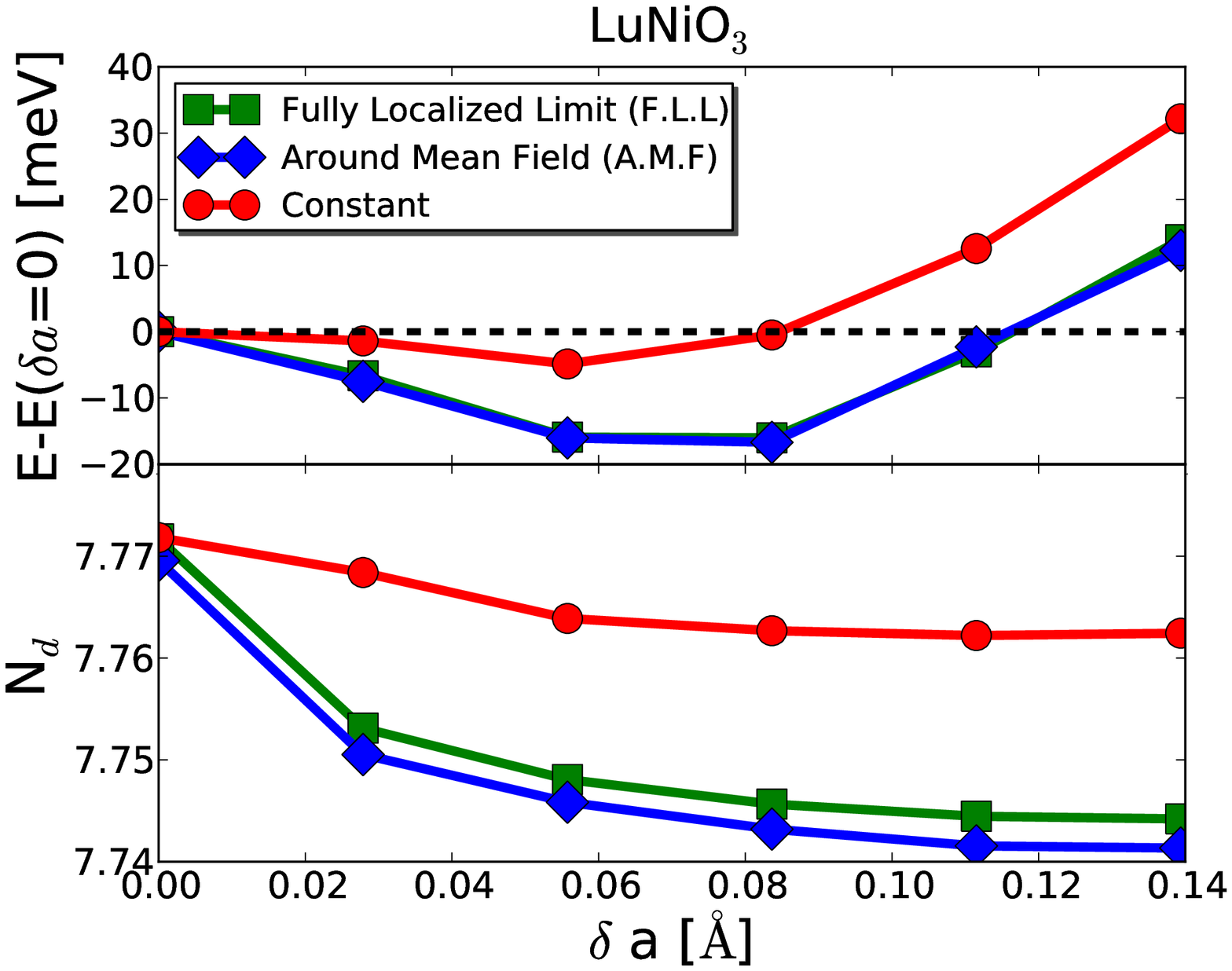}
\includegraphics[scale=0.45]{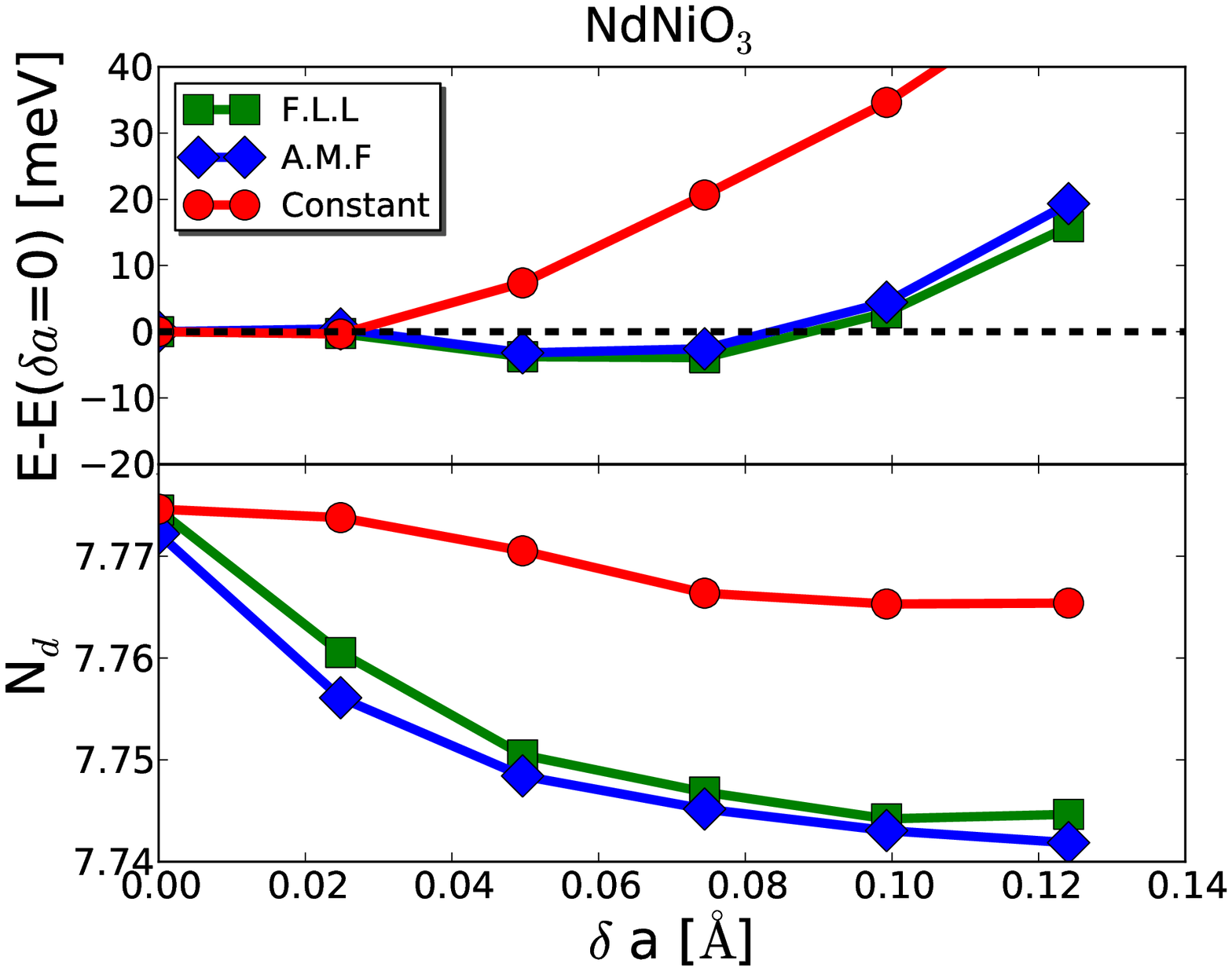}
\caption{(Color online)
Total energy curve $E(\delta a)-E(\delta a=0)$ and the corresponding $N_d$ values along the  $\delta a$ path in LuNiO$_3$ (the zero pressure) and NdNiO$_3$ ($-(V-V_0)/V_0=1.6\%$) computed using  `fully localized limit' (FLL)  (Eq.$\:$\ref{eq:DC}, green square dots), 
the `around mean field' (AMF) limit formula (Eq.$\:$\ref{eq:DC_AMF}, blue diamond dots), and  constant double counting potential (Eq.$\:$\ref{eq:DC4}, red circle dots). The calculations are calibrated such that $N_d=7.775$ for NdNiO$_3$ and $N_d=7.770$ for LuNiO$_3$ in the undistorted structure. The corresponding $U'$ are $4.8eV$ for FLL and  $4.79eV$ for AMF (for both La and Lu), while $\alpha^{dc}$ is 17.97eV for LuNiO$_3$ and 17.98eV for NdNi$O_3$.
\label{fig:double}}
\end{figure}

In this section we compare  results obtained for structural distortions obtained using different double counting formulas: the `fully localized limit' (FLL; Eq.$\:$\ref{eq:DC}), the `around mean field' (AMF; Eq.$\:$\ref{eq:DC_AMF}) and the constant double counting (Eq.$\:$\ref{eq:DC4}). The need to compare predictions for energy changes as a function of structural distortion means we must set up the comparison so that the starting points for the three methods are similar. Because   the physics is controlled by the d-level occupancy, $N_d$, we choose parameters ($U^\prime$ for FLL and AMF; $\alpha^{dc}$ for the constant double counting procedure) in such a way that the $N_d$ for the undistorted structures are the same for all three methods. We choose as our starting point the value $N_d=7.770$ for LuNiO$_3$ and $N_d=7.775$ for NdNiO$_3$ obtained using $U$=5eV and $U^\prime$=4.8eV FLL double counting.
The $U'$ and $\alpha$ values that produce this $N_d$ for the AMF and constant shift double countings are given
in the caption of Fig.$\:$\ref{fig:double}. Then keeping $U$, $U'$, and $\alpha$ fixed we compute the total energy
as a function of distortion $\delta a$ along the path defined previously.

The upper portions of the two panels in Fig.$\:$\ref{fig:double} show the dependence of the total energy on $\delta a$  for LuNiO$_3$ and NdNiO$_3$. The FLL and AMF formulas produce almost identical results for the energy differences (at fixed $U^\prime=U$ the FLL and AMF methods do produce different $N_d$ and different total energies, and as $U$ is varied they predict different locations of the metal-insulator phase boundaries~\cite{Karolak:10} but this is not relevant for the present discussion). The constant double counting potential however gives a significantly different energy curve, predicting in particular a strongly reduced  value of the $\delta a$ that minimize the energy.
For NdNiO$_3$ the constant shift double counting almost completely removes the distortion. 
We suggest that the difference between the constant shift
and the other methodologies arises because the
constant shift formula does not correspond
to an interaction energy term in the Hamiltonian; 
for this reason the contribution to the energy change
arising from correlations may be underestimated. 
Mathematically, because it is just a fixed change in the potential,
the constant shift formula does not allow for a complete treatment of
the feedback between structure and correlation physics
which the other interaction-energy derived formulas incorporate.
To understand one aspect of the differences, we show
in the lower panels of Fig.$\:$\ref{fig:double} the dependence of $N_d$, averaged over the Ni sites, 
as a function of distortion.
The change in $N_d$ is almost the same for the AMF and FLL double countings, 
and for both of these is much greater than for the constant shift double counting.
While these differences in $N_d$ are small, they are relevant on the scale of the
stabilization energy of the distortion.

\section{Conclusion}

In this paper, we have implemented  a fully charge-self-consistent DFT+DMFT method. The method uses the Marzari-Vanderbilt MLWF construction to define the correlated subspace which is treated within DMFT, while the remaining portion of the problem is treated using  a plane-wave basis within the PAW formalism. The combination of a plane wave basis for the density functional calculation and a MLWF representation for the correlated orbitals and those which are hybridized with them enables an efficient formulation and solution of the DFT+DMFT equations, allowing for calculations of   large unit-cells having complex distortions.  The local self-energy of the correlated subspace is obtained using DFMT, and the DMFT  impurity problem is solved using the numerically exact continuous time QMC method~\cite{Werner:06,Werner:2006,Haule:07,Gull_review:11}.The power of the DFT+DMFT method is demonstrated by total energy calculations of  the structural and metal-insulator phase diagrams of the strongly correlated rare-earth nickelates. The experimental phase diagram in the plane of rare earth ion and applied pressure is quantitatively reproduced.   

The DFT+DMFT total energy calculations can correctly capture  the experimental ground-state properties of nickelates in terms of both structural ($Pbnm$ vs $P2_1/n$) and electronic (metal vs insulator) ground states (see Fig.$\:$\ref{fig:phased}). Moreover, the bond-length difference $\delta a$ as a function of volume is quantitatively reproduced (see Fig.$\:$\ref{fig:bondvsvol}). The widely used DFT+U approximation is  implemented using the exact same implementation and found to  grossly overestimate regime of parameter space where  the bond-length disproportionated and insulating phases are found.

We also addressed the importance of choosing a proper double counting potential.  We presented a generalized version of the widely-used FLL and AMF double counting formula, in which the pre-factor $U$ is replaced by a different factor  $U'$  (Eq.$\:$\ref{eq:DC}). This alternative double counting formula can be straightforwardly integrated into the total energy calculations and  produces a consistent phase diagram of nickelates compared with experiment. Different $U'$ values in this double counting formula  change the phase diagram in a significant way (see Fig.$\:$\ref{fig:bondUp}). We argue that the correct value of $U^\prime$ is the one that both reproduces the proper structural energetics and the experimentally observed energy of the oxygen spectra. We found that if $U=5eV$ and $U^\prime=4.8eV$ are chosen both the photo-emission spectra and the energetics are well reproduced, within both the FLL and AMF schemes, for the entire family of nickelates studied in this paper.

All of the calculations presented here are for paramagnetic states. Allowing for static spin polarization raises interesting issues to be addressed  in future work. The questions of whether the DFT portion of the calculation should involve a spin-polarized method such as the local-spin-density-approximation and whether a spin-dependent double counting is needed require further investigation.

Our total energy method can be applied to many systems in which the structural change is closely tied to their electronic transitions, including dimerized VO$_2$ \cite{Biermann05} and actinides with anomalous structural transitions. Studies of phonons and their interactions in correlated materials seem also to be within reach. While we have not yet implemented the computation of forces and stresses in our formalism, recent work \cite{Leonov:14} indicates that this is tractable in the Wannier basis we use.

The results presented in this paper show that DFT+DMFT, although not yet fully $ab$-$initio$ because  values of $U$ for the interactions and $U^\prime$ for the double counting must be determined, is a very promising method for study of the   structural and electronic properties  of complex, strongly correlated electronic systems.  Progress has been made in reliable first-principles approaches to computing $U$~\cite{Aryasetiawan06,Vaugier12}, but more work needs to be done to understand how we might compute $U^\prime$ without experimental input.

\section*{Acknowlegements}

The authors are grateful to Kristjan Haule and Gabriel Kotliar for helpful discussions.  AJM acknowledges funding from the  US Department of Energy under grant DOE-FG02-04-ER046169.  HP and CAM acknowledge funding from FAME, one of six centers of STARnet, a Semiconductor Research Corporation program sponsored by
MARCO and DARPA.

\section*{Appendix A: Local coordinate transformation \label{AppendixA}}

In this paper, we define the correlated subspace and hybridization window using
the Marzari-Vanderbilt MLWF scheme.  The  Wannier functions should provide a good
representations of atomic-like orbitals, including centering the orbital on the ion in addition
to transforming as the
appropriate irreducible representation of the point group when symmetry is present.  
However, often there are small deviations from a symmetry group, and it is desirable
to find the best possible basis which nearly respects the symmetry of the higher group.
For example, many transition metal oxides crystallize in a structure
characterized by a four-sublattice rotation of the transition metal-oxygen
octahedra with respect to the ideal cubic structure. In these circumstances the
Wannier functions representing $d$ electrons on a given transition metal site may
have mixed $e_g$ and $t_{2g}$ character, so that the self energy and
hybridization function have off-diagonal components which introduce a severe
sign problem into quantum Monte Carlo calculations~\cite{Gull_review:11} when
performing DFT+DMFT. It is desirable to avoid this by working with a nearly diagonal
representation of the correlated subspace, which would be some linear combination of
the MLWF which comprise the correlated subspace.
This may be thought of as  aligning the
Wannier basis on a given transition metal ion to the local coordinates
describing the orientation of the relevant oxygen octahedron (although
additional band structure details mean that the optimal local basis is not
exactly aligned to the octahedron).

Therefore, we introduce an additional SO(3) rotational matrix $\hat{\Theta}^{\tau}_{corr}$ acting on the  MLWF in the correlated subspace at each correlated-site $\tau$ within the unit cell such that the Hamiltonian $\hat{H}^{\tau}_{corr}$ is rotated by
\begin{equation}
\hat{H}^{\prime \tau}_{corr} = (\hat{\Theta}^{\tau}_{corr})^{\dagger}(\alpha,\beta,\gamma)\cdot\hat{H}^{\tau}_{corr}\cdot\hat{\Theta}^{\tau}_{corr}(\alpha,\beta,\gamma).
\end{equation}
The Euler angles $\alpha,\beta,\gamma$ at are determined to minimize the sum of the squares of the off-diagonal components in each Hamiltonian $\hat{H}^{\prime \tau}_{corr}$.

The full Hamiltonian in the hybridization window $\hat{H}_{hw}^\prime$ is then given by
\begin{equation}
\hat{H}^{\prime}_{hw} = \hat{\Lambda}^{\dagger}\cdot\hat{H}_{hw}\cdot\hat{\Lambda} 
\label{eq:rot}
\end{equation}
where $\hat{\Lambda}=\hat{\Theta}^I_{corr}\oplus\hat{\Theta}^J_{corr}\oplus\hat{\Theta}^K_{corr}\oplus\cdots\oplus \hat{I}_\ell$, $\tau=I$,$J$,$K$,$\cdots$ are indices of the correlated sites (ie. Ni in our paper) within the unit cell, and  $\ell$ is the dimension of the hybridization window minus the dimension of the correlated subspace.

This additional unitary transform is applied to the maximally localized Wannier functions in Eq.$\:$\ref{eq:Wan}
resulting in the total unitary transform of Eq.$\:$\ref{eq:WanU}. Finally, we obtain the Wannier function in Eq.$\:$\ref{eq:Wan2}  which is used for defining the correlated subspace.

\section*{Appendix B: DMFT self-consistency \label{AppendixB}}

Here we present additional details relevant for DMFT self-consistency of the correlated Green's function. 
A key aspect of the approach is to
define a {\em hybridization window} of states which includes both the correlated
states  (the Ni-$d$ states in the example we consider) and all of the band states
to which they hybridize. In practice we define these via a modified Marzari-Vanderbilt
Maximally Localized Wannier Function construction (see Appendix A for details) applied to the hybridization window
(basically the Ni-$3d$ states and the O-$2p$ states, in the nickelates we consider herein).

By construction the DFT Hamiltonian, denoted as $\hat{H}_0$ in this appendix, is then block diagonal, with no matrix elements mixing states in the hybridization window with states outside it. Expressing the relevant portion of  $H_0$ in the Wannier representation of the hybridization window gives
\begin{equation}
\hat{H}^{0}_{\mathbf{k}mn}=\langle \bar{W}_{\mathbf{k}m}|-\frac{1}{2}\hat{\nabla}^{2}+\hat{V}^{ext}+\hat{V}^{Hxc}|\bar{W}_{\mathbf{k}n}\rangle
\label{eq:HDFT}
\end{equation}
where $m$, $n$ are  dual indices $(\tau,\alpha)$ in which $\tau$ labels an atom in the unit cell and $\alpha$ labels the orbital character of the corresponding site, and 
$\bar{W}_{\mathbf{k}m}$ are the Fourier transform in the first Brillouin zone of the functions defined in Eq.~\ref{eq:Wan2}.

The Green's function  $\hat{G}$ in Eq.$\:$\ref{Gdef} is similarly block diagonal; the portion acting on the hybridization window is  obtained by inverting the operator 
\begin{equation}
\hat{G}^{hw}_{\mathbf{k}}(i\omega_n)=\left[ i\omega_n\mathbf{1}+\mu-\hat{H}^{0}_{\mathbf{k}}-\hat{P}_{cor}^\dagger(\hat{\Sigma}_{loc}(i\omega_n)-\hat{V}^{DC})\hat{P}_{cor} \right]^{-1}
\label{Ginveq}
\end{equation}
where $\hat{\Sigma}$ and $\hat{V}^{DC}$ are operators with non-zero matrix elements only in the correlated subspace of the hybridization window.

The effective Hamiltonian defined by $i\omega_n\mathbf{1}-\hat{G}^{hw}_{\mathbf{k}}(i\omega_n)$ is non-Hermitian because the self energy on the diagonal component is complex.  Its  eigenvalues are complex numbers and its left and right eigenvectors are generally not complex conjugate to each other. Inversion is accomplished through the solution of the  generalized eigenvalue equation
\begin{equation}
\left[\hat{H}^{0}_{\mathbf{k}} +\hat{P}_{cor}^\dagger(\hat{\Sigma}_{loc}(i\omega_n)-\hat{V}^{DC})\hat{P}_{cor} \right]
|C_{\mathbf{k}l}^{R,i\omega}\rangle = \epsilon_{\mathbf{k}l}^{\omega_n}
|C_{\mathbf{k}l}^{R,i\omega}\rangle
\label{eq:1}
\end{equation}
with $\epsilon$ a the complex eigenvalue and $C^{R}$  the right eigenfunctions.
The Green's function of the hybridization window can then be represented in terms of the frequency dependent eigenvalues and left/right eigenfunctions obtained from Eq.$\:$\ref{eq:1} as
\begin{eqnarray}
\hat{G}^{hw}_{\mathbf{k}}(i\omega_n) & = & \sum_l
\frac{|C_{\mathbf{k}l}^{R,i\omega}\rangle  \langle C_{\mathbf{k}l}^{L,i\omega}|
}{i\omega_{n}+\mu-\epsilon_{\mathbf{k}l}^{\omega_{n}}}.
\label{eq:2}
\end{eqnarray}

Once the Green's function $\hat{G}^{hw}_{\mathbf{k}}(i\omega_n)$ is obtained, the DMFT
self consistency condition requires that the impurity model Green's function
$\hat{G}^{imp}$, a matrix with dimension of the correlated subspace, is given by the local projection of
$\hat{G}^{hw}_{\mathbf{k}}(i\omega_n)$ into the correlated subspace; thus
$\hat{G}^{imp}=
\frac{1}{N_{\mathbf{k}}}\sum_{\mathbf{k}} \hat{P}_{cor}  \hat{G}^{hw}_{\mathbf{k}}(i\omega_n) \hat{P}_{cor}^\dagger $.
The hybridization function $\Delta(i\omega_n)$ for the  auxiliary impurity 
is given by
\begin{eqnarray}
\hat{\Delta}(i\omega_n) & = & (i\omega+\mu)\cdot\hat{\mathbb{I}}-\hat{\epsilon}_{imp}-\hat{\Sigma}_{loc}(i\omega_n)
\\
&&-\left[\frac{1}{N_{\mathbf{k}}}\sum_{\mathbf{k}}   \hat{P}_{cor}  \hat{G}^{hw}_{\mathbf{k}}(i\omega_n) \hat{P}_{cor}^\dagger \right]^{-1}
\end{eqnarray}
where $\hat{\epsilon}_{imp}$ is the impurity level matrix.

Using this new $\hat{\Delta}(i\omega_n)$, the new self energy $\hat{\Sigma}^{imp}(i\omega_n)$ is obtained from the quantum impurity solver
and identified as $\hat{\Sigma}_{loc}(i\omega_n)$,
the new $\hat{G}^{hw}_{\mathbf{k}}(i\omega_n)$ is then constructed, and then the entire process is repeated until convergence is achieved.
In practice, we determine the convergence by monitoring the correlation energy part, $E^{pot}-E^{DC}$ (see Eq.$\:$\ref{Epot} and Section \ref{sec:Theory-5}); the DMFT loop is converged if the change in $E^{pot}-E^{DC}$ is less than 1meV from one iteration to the next.

The chemical potential $\mu$ is determined  such that the total number of electrons within the hybridization window is equal to the appropriate integer for the system at hand,
which would be 25 per formula unit for the nickelates in this study. 
For numerical accuracy it is advantageous to treat the high frequency tail of $\hat{G}^{hw}_{\mathbf{k}}(i\omega_n)$ analytically, by noting that at high energies the self energy vanishes so that 
\begin{eqnarray}
N_{tot} & = & \frac{T}{N_{\mathbf{k}}}\sum_{\mathbf{k},j,\omega_n}G^{hw}_{\mathbf{k}jj}(i\omega_n)
\label{eq:3}\\
&=&\frac{1}{N_{\mathbf{k}}}\sum_{\mathbf{k},l}f(\epsilon_{\mathbf{k}l}^{\omega_{\infty}}-\mu)+\frac{T}{N_{\mathbf{k}}}\nonumber \\
&&\sum_{\mathbf{k},\omega_n,l}\left(\frac{1}{i\omega_n
+\mu-\epsilon_{\mathbf{k}l}^{\omega_{n}}}-\frac{1}{i\omega_n+\mu-\epsilon_{\mathbf{k}l}^{\omega_{\infty}}}\right)
\nonumber
\end{eqnarray}
where $f(\epsilon)$ is the Fermi function and $\epsilon_{\mathbf{k}l}^{\omega_{\infty}}$
is the eigenvalue of Eq.$\:$\ref{eq:1} evaluated at $\omega\rightarrow\infty$.

The density matrix at each momentum $\mathbf{k}$ and orbital indices $m$, $n$ is obtained by taking a trace of $\hat{G}^{hw}_{\mathbf{k}}(i\omega_n)$ over only frequency and in analogy to Eq.~\ref{eq:3} is 
\begin{eqnarray}
\hat{n}_{\mathbf{k}} & = & T\cdot \sum_{i\omega_n}\hat{G}^{hw}_{\mathbf{k}}(i\omega_n)  \nonumber \\
& = & T\cdot\sum_{l,\omega_{n}}\left(  
\frac{ 
|C_{\mathbf{k}l}^{R,i\omega}\rangle  \langle C_{\mathbf{k}l}^{L,i\omega}| 
}
{i\omega_{n}+\mu-\epsilon_{\mathbf{k}l}^{\omega_{n}}}-
\frac{
|C_{\mathbf{k}l}^{\omega_{\infty}}\rangle  \langle C_{\mathbf{k}l}^{\omega_{\infty}}| 
}{i\omega_{n}
+\mu-\epsilon_{\mathbf{k}l}^{\omega_{\infty}}}\right) \nonumber \\
& & +
\sum_{l}
|C_{\mathbf{k}l}^{\omega_{\infty}}\rangle  \langle C_{\mathbf{k}l}^{\omega_{\infty}}|
f(\epsilon_{\mathbf{k}l}^{\omega_{\infty}}-\mu)
\label{eq:4}
\end{eqnarray}
where $\epsilon_{\mathbf{k}l}^{\omega_{\infty}}$ and $C_{\textbf{k}ml}^{\omega_{\infty}}$
are the solutions of the eigenvalue problem in Eq.$\:$\ref{eq:1} at $i\omega_{n}\rightarrow\infty$.
The $d$ orbital occupancy $N_{d}$ is defined as the trace of the density matrix within the manifold of correlated states. 
\begin{eqnarray}
N_{d} & = & \frac{1}{N_{\mathbf{k}}}\sum_{\mathbf{k}m}n_{\mathbf{k}}^{(m,m)}
\label{eq:5}
\end{eqnarray}

\section*{Appendix C: The Hubbard $U$ dependence on the phase transition}

\begin{figure}[!htbp]
\includegraphics[scale=0.45]{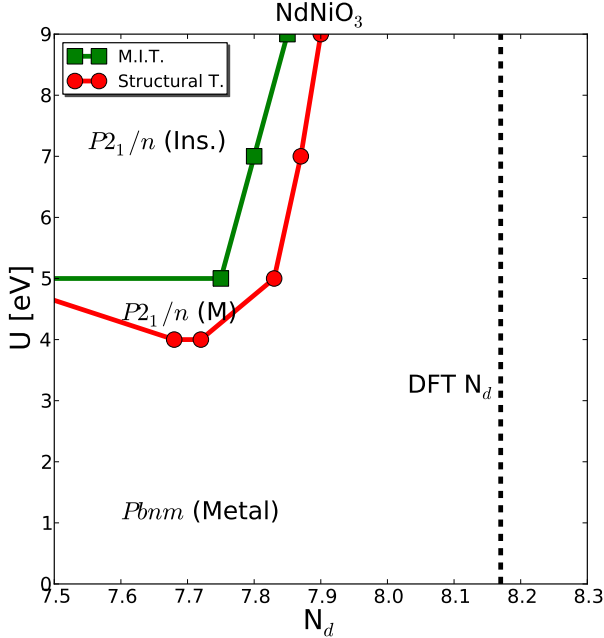}
\caption{(Color online)
Diagram showing the location of the metal-insulator transition (squares, green on line)
and  structural transition (circles, red on line)  of NdNiO$_3$ at
$-(V-V_0)/V_0=1.6\%$ determined as described in the main text, as a function of
the Hubbard interaction $U$ (y-axis) and the $d$-occupancy $N_d$ (x-axis).
Above and to the left of the lines the system is predicted to be distorted and
insulating; below and to the right, undistorted and metallic. 
\label{fig:U_Nd}}
\end{figure}

In this Appendix, we show the dependence of the metal-insulator and structural
phase diagrams of rare-earth nickelates on the magnitude of the on-site Hubbard
interaction $U$. Fig.$\:$\ref{fig:U_Nd} displays the DFT+DMFT phase diagram of
NdNiO$_3$ in the plane of Hubbard interaction $U$ (y-axis) and the
$d$-occupancy $N_d$ (x-axis). The $d$-occupancy parametrizes the 
energy difference between Ni $d$ and O $p$ orbitals, which in
turn is controlled by the double counting parameter $U'$. Therefore, each
 $N_d$ point on the x-axis corresponds to a given $U'$.

At $U$=0, both the structural (circle dots) and the metal-insulator (square
dots) transitions do not occur at any $N_d$ value: the transition is a
correlation effect.  Above a threshold $U$  ($\sim$4eV) a
bond-disproportionated, insulating  $P2_1/n$ structure occurs {\em if}  $N_d$
is small enough, but if $N_d$ is too large, even a very large $U$ will not
drive a structural or metal-insulator transition. We also note that the $N_d$
predicted by DFT calculations is far from the value required to drive the
transition, again indicating the importance of  an appropriate double counting.
The phase diagram in Fig.$\:$\ref{fig:U_Nd} provides an important additional
perspective on the importance of fixing the double counting correction in
correlated materials.

\bibliography{main}

\begin{thebibliography}{70}%
\makeatletter
\providecommand \@ifxundefined [1]{%
 \@ifx{#1\undefined}
}%
\providecommand \@ifnum [1]{%
 \ifnum #1\expandafter \@firstoftwo
 \else \expandafter \@secondoftwo
 \fi
}%
\providecommand \@ifx [1]{%
 \ifx #1\expandafter \@firstoftwo
 \else \expandafter \@secondoftwo
 \fi
}%
\providecommand \natexlab [1]{#1}%
\providecommand \enquote  [1]{``#1''}%
\providecommand \bibnamefont  [1]{#1}%
\providecommand \bibfnamefont [1]{#1}%
\providecommand \citenamefont [1]{#1}%
\providecommand \href@noop [0]{\@secondoftwo}%
\providecommand \href [0]{\begingroup \@sanitize@url \@href}%
\providecommand \@href[1]{\@@startlink{#1}\@@href}%
\providecommand \@@href[1]{\endgroup#1\@@endlink}%
\providecommand \@sanitize@url [0]{\catcode `\\12\catcode `\$12\catcode
  `\&12\catcode `\#12\catcode `\^12\catcode `\_12\catcode `\%12\relax}%
\providecommand \@@startlink[1]{}%
\providecommand \@@endlink[0]{}%
\providecommand \url  [0]{\begingroup\@sanitize@url \@url }%
\providecommand \@url [1]{\endgroup\@href {#1}{\urlprefix }}%
\providecommand \urlprefix  [0]{URL }%
\providecommand \Eprint [0]{\href }%
\providecommand \doibase [0]{http://dx.doi.org/}%
\providecommand \selectlanguage [0]{\@gobble}%
\providecommand \bibinfo  [0]{\@secondoftwo}%
\providecommand \bibfield  [0]{\@secondoftwo}%
\providecommand \translation [1]{[#1]}%
\providecommand \BibitemOpen [0]{}%
\providecommand \bibitemStop [0]{}%
\providecommand \bibitemNoStop [0]{.\EOS\space}%
\providecommand \EOS [0]{\spacefactor3000\relax}%
\providecommand \BibitemShut  [1]{\csname bibitem#1\endcsname}%
\let\auto@bib@innerbib\@empty
%</preamble>
\bibitem [{\citenamefont {Kotliar}\ \emph {et~al.}(2006)\citenamefont
  {Kotliar}, \citenamefont {Savrasov}, \citenamefont {Haule}, \citenamefont
  {Oudovenko}, \citenamefont {Parcollet},\ and\ \citenamefont
  {Marianetti}}]{Kotliar:06}%
  \BibitemOpen
  \bibfield  {author} {\bibinfo {author} {\bibfnamefont {G.}~\bibnamefont
  {Kotliar}}, \bibinfo {author} {\bibfnamefont {S.~Y.}\ \bibnamefont
  {Savrasov}}, \bibinfo {author} {\bibfnamefont {K.}~\bibnamefont {Haule}},
  \bibinfo {author} {\bibfnamefont {V.~S.}\ \bibnamefont {Oudovenko}}, \bibinfo
  {author} {\bibfnamefont {O.}~\bibnamefont {Parcollet}}, \ and\ \bibinfo
  {author} {\bibfnamefont {C.~A.}\ \bibnamefont {Marianetti}},\ }\href
  {\doibase 10.1103/RevModPhys.78.865} {\bibfield  {journal} {\bibinfo
  {journal} {Rev. Mod. Phys.}\ }\textbf {\bibinfo {volume} {78}},\ \bibinfo
  {pages} {865} (\bibinfo {year} {2006})}\BibitemShut {NoStop}%
\bibitem [{\citenamefont {Savrasov}\ \emph {et~al.}(2001)\citenamefont
  {Savrasov}, \citenamefont {Kotliar},\ and\ \citenamefont
  {Abrahams}}]{Savrasov:01}%
  \BibitemOpen
  \bibfield  {author} {\bibinfo {author} {\bibfnamefont {S.~Y.}\ \bibnamefont
  {Savrasov}}, \bibinfo {author} {\bibfnamefont {G.}~\bibnamefont {Kotliar}}, \
  and\ \bibinfo {author} {\bibfnamefont {E.}~\bibnamefont {Abrahams}},\
  }\href@noop {} {\bibfield  {journal} {\bibinfo  {journal} {Nature}\ }\textbf
  {\bibinfo {volume} {410}},\ \bibinfo {pages} {793} (\bibinfo {year}
  {2001})}\BibitemShut {NoStop}%
\bibitem [{\citenamefont {Andersen}(1975)}]{Anderson:75}%
  \BibitemOpen
  \bibfield  {author} {\bibinfo {author} {\bibfnamefont {O.~K.}\ \bibnamefont
  {Andersen}},\ }\href {\doibase 10.1103/PhysRevB.12.3060} {\bibfield
  {journal} {\bibinfo  {journal} {Phys. Rev. B}\ }\textbf {\bibinfo {volume}
  {12}},\ \bibinfo {pages} {3060} (\bibinfo {year} {1975})}\BibitemShut
  {NoStop}%
\bibitem [{\citenamefont {Savrasov}\ \emph {et~al.}(2005)\citenamefont
  {Savrasov}, \citenamefont {Oudovenko}, \citenamefont {Haule}, \citenamefont
  {Villani},\ and\ \citenamefont {Kotliar}}]{Savrasov2005115117}%
  \BibitemOpen
  \bibfield  {author} {\bibinfo {author} {\bibfnamefont {S.~Y.}\ \bibnamefont
  {Savrasov}}, \bibinfo {author} {\bibfnamefont {V.}~\bibnamefont {Oudovenko}},
  \bibinfo {author} {\bibfnamefont {K.}~\bibnamefont {Haule}}, \bibinfo
  {author} {\bibfnamefont {D.}~\bibnamefont {Villani}}, \ and\ \bibinfo
  {author} {\bibfnamefont {G.}~\bibnamefont {Kotliar}},\ }\href {\doibase
  10.1103/PhysRevB.71.115117} {\bibfield  {journal} {\bibinfo  {journal} {Phys.
  Rev. B}\ }\textbf {\bibinfo {volume} {71}},\ \bibinfo {pages} {115117}
  (\bibinfo {year} {2005})}\BibitemShut {NoStop}%
\bibitem [{\citenamefont {Held}\ \emph {et~al.}(2001)\citenamefont {Held},
  \citenamefont {McMahan},\ and\ \citenamefont {Scalettar}}]{Held:01}%
  \BibitemOpen
  \bibfield  {author} {\bibinfo {author} {\bibfnamefont {K.}~\bibnamefont
  {Held}}, \bibinfo {author} {\bibfnamefont {A.~K.}\ \bibnamefont {McMahan}}, \
  and\ \bibinfo {author} {\bibfnamefont {R.~T.}\ \bibnamefont {Scalettar}},\
  }\href {\doibase 10.1103/PhysRevLett.87.276404} {\bibfield  {journal}
  {\bibinfo  {journal} {Phys. Rev. Lett.}\ }\textbf {\bibinfo {volume} {87}},\
  \bibinfo {pages} {276404} (\bibinfo {year} {2001})}\BibitemShut {NoStop}%
\bibitem [{\citenamefont {McMahan}\ \emph {et~al.}(2003)\citenamefont
  {McMahan}, \citenamefont {Held},\ and\ \citenamefont
  {Scalettar}}]{McMahan:03}%
  \BibitemOpen
  \bibfield  {author} {\bibinfo {author} {\bibfnamefont {A.~K.}\ \bibnamefont
  {McMahan}}, \bibinfo {author} {\bibfnamefont {K.}~\bibnamefont {Held}}, \
  and\ \bibinfo {author} {\bibfnamefont {R.~T.}\ \bibnamefont {Scalettar}},\
  }\href {\doibase 10.1103/PhysRevB.67.075108} {\bibfield  {journal} {\bibinfo
  {journal} {Phys. Rev. B}\ }\textbf {\bibinfo {volume} {67}},\ \bibinfo
  {pages} {075108} (\bibinfo {year} {2003})}\BibitemShut {NoStop}%
\bibitem [{\citenamefont {Amadon}\ \emph {et~al.}(2006)\citenamefont {Amadon},
  \citenamefont {Biermann}, \citenamefont {Georges},\ and\ \citenamefont
  {Aryasetiawan}}]{Amadon:06}%
  \BibitemOpen
  \bibfield  {author} {\bibinfo {author} {\bibfnamefont {B.}~\bibnamefont
  {Amadon}}, \bibinfo {author} {\bibfnamefont {S.}~\bibnamefont {Biermann}},
  \bibinfo {author} {\bibfnamefont {A.}~\bibnamefont {Georges}}, \ and\
  \bibinfo {author} {\bibfnamefont {F.}~\bibnamefont {Aryasetiawan}},\ }\href
  {\doibase 10.1103/PhysRevLett.96.066402} {\bibfield  {journal} {\bibinfo
  {journal} {Phys. Rev. Lett.}\ }\textbf {\bibinfo {volume} {96}},\ \bibinfo
  {pages} {066402} (\bibinfo {year} {2006})}\BibitemShut {NoStop}%
\bibitem [{\citenamefont {Leonov}\ \emph {et~al.}(2010)\citenamefont {Leonov},
  \citenamefont {Korotin}, \citenamefont {Binggeli}, \citenamefont {Anisimov},\
  and\ \citenamefont {Vollhardt}}]{Leonov:10}%
  \BibitemOpen
  \bibfield  {author} {\bibinfo {author} {\bibfnamefont {I.}~\bibnamefont
  {Leonov}}, \bibinfo {author} {\bibfnamefont {D.}~\bibnamefont {Korotin}},
  \bibinfo {author} {\bibfnamefont {N.}~\bibnamefont {Binggeli}}, \bibinfo
  {author} {\bibfnamefont {V.~I.}\ \bibnamefont {Anisimov}}, \ and\ \bibinfo
  {author} {\bibfnamefont {D.}~\bibnamefont {Vollhardt}},\ }\href {\doibase
  10.1103/PhysRevB.81.075109} {\bibfield  {journal} {\bibinfo  {journal} {Phys.
  Rev. B}\ }\textbf {\bibinfo {volume} {81}},\ \bibinfo {pages} {075109}
  (\bibinfo {year} {2010})}\BibitemShut {NoStop}%
\bibitem [{\citenamefont {Leonov}\ \emph {et~al.}(2008)\citenamefont {Leonov},
  \citenamefont {Binggeli}, \citenamefont {Korotin}, \citenamefont {Anisimov},
  \citenamefont {Stoji\ifmmode~\acute{c}\else \'{c}\fi{}},\ and\ \citenamefont
  {Vollhardt}}]{Leonov:08}%
  \BibitemOpen
  \bibfield  {author} {\bibinfo {author} {\bibfnamefont {I.}~\bibnamefont
  {Leonov}}, \bibinfo {author} {\bibfnamefont {N.}~\bibnamefont {Binggeli}},
  \bibinfo {author} {\bibfnamefont {D.}~\bibnamefont {Korotin}}, \bibinfo
  {author} {\bibfnamefont {V.~I.}\ \bibnamefont {Anisimov}}, \bibinfo {author}
  {\bibfnamefont {N.}~\bibnamefont {Stoji\ifmmode~\acute{c}\else \'{c}\fi{}}},
  \ and\ \bibinfo {author} {\bibfnamefont {D.}~\bibnamefont {Vollhardt}},\
  }\href {\doibase 10.1103/PhysRevLett.101.096405} {\bibfield  {journal}
  {\bibinfo  {journal} {Phys. Rev. Lett.}\ }\textbf {\bibinfo {volume} {101}},\
  \bibinfo {pages} {096405} (\bibinfo {year} {2008})}\BibitemShut {NoStop}%
\bibitem [{\citenamefont {Leonov}\ \emph {et~al.}(2011)\citenamefont {Leonov},
  \citenamefont {Poteryaev}, \citenamefont {Anisimov},\ and\ \citenamefont
  {Vollhardt}}]{Leonov:11}%
  \BibitemOpen
  \bibfield  {author} {\bibinfo {author} {\bibfnamefont {I.}~\bibnamefont
  {Leonov}}, \bibinfo {author} {\bibfnamefont {A.~I.}\ \bibnamefont
  {Poteryaev}}, \bibinfo {author} {\bibfnamefont {V.~I.}\ \bibnamefont
  {Anisimov}}, \ and\ \bibinfo {author} {\bibfnamefont {D.}~\bibnamefont
  {Vollhardt}},\ }\href {\doibase 10.1103/PhysRevLett.106.106405} {\bibfield
  {journal} {\bibinfo  {journal} {Phys. Rev. Lett.}\ }\textbf {\bibinfo
  {volume} {106}},\ \bibinfo {pages} {106405} (\bibinfo {year}
  {2011})}\BibitemShut {NoStop}%
\bibitem [{\citenamefont {Hubbard}(1963)}]{Hubbard26111963}%
  \BibitemOpen
  \bibfield  {author} {\bibinfo {author} {\bibfnamefont {J.}~\bibnamefont
  {Hubbard}},\ }\href {\doibase 10.1098/rspa.1963.0204} {\bibfield  {journal}
  {\bibinfo  {journal} {Proc. R. Soc. London, Ser. A.}\ }\textbf {\bibinfo
  {volume} {276}},\ \bibinfo {pages} {238} (\bibinfo {year}
  {1963})}\BibitemShut {NoStop}%
\bibitem [{\citenamefont {Pourovskii}\ \emph {et~al.}(2007)\citenamefont
  {Pourovskii}, \citenamefont {Amadon}, \citenamefont {Biermann},\ and\
  \citenamefont {Georges}}]{Pourovskii:07}%
  \BibitemOpen
  \bibfield  {author} {\bibinfo {author} {\bibfnamefont {L.~V.}\ \bibnamefont
  {Pourovskii}}, \bibinfo {author} {\bibfnamefont {B.}~\bibnamefont {Amadon}},
  \bibinfo {author} {\bibfnamefont {S.}~\bibnamefont {Biermann}}, \ and\
  \bibinfo {author} {\bibfnamefont {A.}~\bibnamefont {Georges}},\ }\href
  {\doibase 10.1103/PhysRevB.76.235101} {\bibfield  {journal} {\bibinfo
  {journal} {Phys. Rev. B}\ }\textbf {\bibinfo {volume} {76}},\ \bibinfo
  {pages} {235101} (\bibinfo {year} {2007})}\BibitemShut {NoStop}%
\bibitem [{\citenamefont {Amadon}(2012)}]{Amadon:12}%
  \BibitemOpen
  \bibfield  {author} {\bibinfo {author} {\bibfnamefont {B.}~\bibnamefont
  {Amadon}},\ }\href@noop {} {\bibfield  {journal} {\bibinfo  {journal}
  {Journal of Physics: Condensed Matter}\ }\textbf {\bibinfo {volume} {24}},\
  \bibinfo {pages} {075604} (\bibinfo {year} {2012})}\BibitemShut {NoStop}%
\bibitem [{\citenamefont {Di~Marco}\ \emph {et~al.}(2009)\citenamefont
  {Di~Marco}, \citenamefont {Min\'ar}, \citenamefont {Chadov}, \citenamefont
  {Katsnelson}, \citenamefont {Ebert},\ and\ \citenamefont
  {Lichtenstein}}]{Marco:09}%
  \BibitemOpen
  \bibfield  {author} {\bibinfo {author} {\bibfnamefont {I.}~\bibnamefont
  {Di~Marco}}, \bibinfo {author} {\bibfnamefont {J.}~\bibnamefont {Min\'ar}},
  \bibinfo {author} {\bibfnamefont {S.}~\bibnamefont {Chadov}}, \bibinfo
  {author} {\bibfnamefont {M.~I.}\ \bibnamefont {Katsnelson}}, \bibinfo
  {author} {\bibfnamefont {H.}~\bibnamefont {Ebert}}, \ and\ \bibinfo {author}
  {\bibfnamefont {A.~I.}\ \bibnamefont {Lichtenstein}},\ }\href {\doibase
  10.1103/PhysRevB.79.115111} {\bibfield  {journal} {\bibinfo  {journal} {Phys.
  Rev. B}\ }\textbf {\bibinfo {volume} {79}},\ \bibinfo {pages} {115111}
  (\bibinfo {year} {2009})}\BibitemShut {NoStop}%
\bibitem [{\citenamefont {Pourovskii}\ \emph {et~al.}(2005)\citenamefont
  {Pourovskii}, \citenamefont {Katsnelson},\ and\ \citenamefont
  {Lichtenstein}}]{Lichtenstein:05}%
  \BibitemOpen
  \bibfield  {author} {\bibinfo {author} {\bibfnamefont {L.~V.}\ \bibnamefont
  {Pourovskii}}, \bibinfo {author} {\bibfnamefont {M.~I.}\ \bibnamefont
  {Katsnelson}}, \ and\ \bibinfo {author} {\bibfnamefont {A.~I.}\ \bibnamefont
  {Lichtenstein}},\ }\href {\doibase 10.1103/PhysRevB.72.115106} {\bibfield
  {journal} {\bibinfo  {journal} {Phys. Rev. B}\ }\textbf {\bibinfo {volume}
  {72}},\ \bibinfo {pages} {115106} (\bibinfo {year} {2005})}\BibitemShut
  {NoStop}%
\bibitem [{\citenamefont {Werner}\ \emph {et~al.}(2006)\citenamefont {Werner},
  \citenamefont {Comanac}, \citenamefont {deMedici}, \citenamefont {Troyer},\
  and\ \citenamefont {Millis}}]{Werner:06}%
  \BibitemOpen
  \bibfield  {author} {\bibinfo {author} {\bibfnamefont {P.}~\bibnamefont
  {Werner}}, \bibinfo {author} {\bibfnamefont {A.}~\bibnamefont {Comanac}},
  \bibinfo {author} {\bibfnamefont {L.}~\bibnamefont {deMedici}}, \bibinfo
  {author} {\bibfnamefont {M.}~\bibnamefont {Troyer}}, \ and\ \bibinfo {author}
  {\bibfnamefont {A.~J.}\ \bibnamefont {Millis}},\ }\href@noop {} {\bibfield
  {journal} {\bibinfo  {journal} {Phys. Rev. Lett.}\ }\textbf {\bibinfo
  {volume} {97}},\ \bibinfo {pages} {076405} (\bibinfo {year}
  {2006})}\BibitemShut {NoStop}%
\bibitem [{\citenamefont {Werner}\ and\ \citenamefont
  {Millis}(2006)}]{Werner:2006}%
  \BibitemOpen
  \bibfield  {author} {\bibinfo {author} {\bibfnamefont {P.}~\bibnamefont
  {Werner}}\ and\ \bibinfo {author} {\bibfnamefont {A.~J.}\ \bibnamefont
  {Millis}},\ }\href {\doibase 10.1103/PhysRevB.74.155107} {\bibfield
  {journal} {\bibinfo  {journal} {Phys. Rev. B}\ }\textbf {\bibinfo {volume}
  {74}},\ \bibinfo {pages} {155107} (\bibinfo {year} {2006})}\BibitemShut
  {NoStop}%
\bibitem [{\citenamefont {Haule}(2007)}]{Haule:07}%
  \BibitemOpen
  \bibfield  {author} {\bibinfo {author} {\bibfnamefont {K.}~\bibnamefont
  {Haule}},\ }\href {\doibase 10.1103/PhysRevB.75.155113} {\bibfield  {journal}
  {\bibinfo  {journal} {Phys. Rev. B}\ }\textbf {\bibinfo {volume} {75}},\
  \bibinfo {pages} {155113} (\bibinfo {year} {2007})}\BibitemShut {NoStop}%
\bibitem [{\citenamefont {Gull}\ \emph {et~al.}(2011)\citenamefont {Gull},
  \citenamefont {Millis}, \citenamefont {Lichtenstein}, \citenamefont
  {Rubtsov}, \citenamefont {Troyer},\ and\ \citenamefont
  {Werner}}]{Gull_review:11}%
  \BibitemOpen
  \bibfield  {author} {\bibinfo {author} {\bibfnamefont {E.}~\bibnamefont
  {Gull}}, \bibinfo {author} {\bibfnamefont {A.~J.}\ \bibnamefont {Millis}},
  \bibinfo {author} {\bibfnamefont {A.~I.}\ \bibnamefont {Lichtenstein}},
  \bibinfo {author} {\bibfnamefont {A.~N.}\ \bibnamefont {Rubtsov}}, \bibinfo
  {author} {\bibfnamefont {M.}~\bibnamefont {Troyer}}, \ and\ \bibinfo {author}
  {\bibfnamefont {P.}~\bibnamefont {Werner}},\ }\href {\doibase
  10.1103/RevModPhys.83.349} {\bibfield  {journal} {\bibinfo  {journal} {Rev.
  Mod. Phys.}\ }\textbf {\bibinfo {volume} {83}},\ \bibinfo {pages} {349}
  (\bibinfo {year} {2011})}\BibitemShut {NoStop}%
\bibitem [{\citenamefont {Singh}(1994)}]{Singh:94}%
  \BibitemOpen
  \bibfield  {author} {\bibinfo {author} {\bibfnamefont {D.~J.}\ \bibnamefont
  {Singh}},\ }\href@noop {} {\emph {\bibinfo {title} {{Planewaves,
  Pseudopotentials, and the LAPW Method}}}}\ (\bibinfo  {publisher} {Kluwer},\
  \bibinfo {address} {Boston},\ \bibinfo {year} {1994})\BibitemShut {NoStop}%
\bibitem [{\citenamefont {Aichhorn}\ \emph {et~al.}(2011)\citenamefont
  {Aichhorn}, \citenamefont {Pourovskii},\ and\ \citenamefont
  {Georges}}]{Aichhorn:11}%
  \BibitemOpen
  \bibfield  {author} {\bibinfo {author} {\bibfnamefont {M.}~\bibnamefont
  {Aichhorn}}, \bibinfo {author} {\bibfnamefont {L.}~\bibnamefont
  {Pourovskii}}, \ and\ \bibinfo {author} {\bibfnamefont {A.}~\bibnamefont
  {Georges}},\ }\href {\doibase 10.1103/PhysRevB.84.054529} {\bibfield
  {journal} {\bibinfo  {journal} {Phys. Rev. B}\ }\textbf {\bibinfo {volume}
  {84}},\ \bibinfo {pages} {054529} (\bibinfo {year} {2011})}\BibitemShut
  {NoStop}%
\bibitem [{\citenamefont {Lee}\ \emph {et~al.}(2012)\citenamefont {Lee},
  \citenamefont {Ji}, \citenamefont {Kim}, \citenamefont {Kim}, \citenamefont
  {Haule}, \citenamefont {Kotliar}, \citenamefont {Lee}, \citenamefont {Khim},
  \citenamefont {Kim}, \citenamefont {Kim}, \citenamefont {Kim},\ and\
  \citenamefont {Shim}}]{Lee:12}%
  \BibitemOpen
  \bibfield  {author} {\bibinfo {author} {\bibfnamefont {G.}~\bibnamefont
  {Lee}}, \bibinfo {author} {\bibfnamefont {H.~S.}\ \bibnamefont {Ji}},
  \bibinfo {author} {\bibfnamefont {Y.}~\bibnamefont {Kim}}, \bibinfo {author}
  {\bibfnamefont {C.}~\bibnamefont {Kim}}, \bibinfo {author} {\bibfnamefont
  {K.}~\bibnamefont {Haule}}, \bibinfo {author} {\bibfnamefont
  {G.}~\bibnamefont {Kotliar}}, \bibinfo {author} {\bibfnamefont
  {B.}~\bibnamefont {Lee}}, \bibinfo {author} {\bibfnamefont {S.}~\bibnamefont
  {Khim}}, \bibinfo {author} {\bibfnamefont {K.~H.}\ \bibnamefont {Kim}},
  \bibinfo {author} {\bibfnamefont {K.~S.}\ \bibnamefont {Kim}}, \bibinfo
  {author} {\bibfnamefont {K.-S.}\ \bibnamefont {Kim}}, \ and\ \bibinfo
  {author} {\bibfnamefont {J.~H.}\ \bibnamefont {Shim}},\ }\href {\doibase
  10.1103/PhysRevLett.109.177001} {\bibfield  {journal} {\bibinfo  {journal}
  {Phys. Rev. Lett.}\ }\textbf {\bibinfo {volume} {109}},\ \bibinfo {pages}
  {177001} (\bibinfo {year} {2012})}\BibitemShut {NoStop}%
\bibitem [{\citenamefont {Grieger}\ \emph {et~al.}(2012)\citenamefont
  {Grieger}, \citenamefont {Piefke}, \citenamefont {Peil},\ and\ \citenamefont
  {Lechermann}}]{Lechermann:12}%
  \BibitemOpen
  \bibfield  {author} {\bibinfo {author} {\bibfnamefont {D.}~\bibnamefont
  {Grieger}}, \bibinfo {author} {\bibfnamefont {C.}~\bibnamefont {Piefke}},
  \bibinfo {author} {\bibfnamefont {O.~E.}\ \bibnamefont {Peil}}, \ and\
  \bibinfo {author} {\bibfnamefont {F.}~\bibnamefont {Lechermann}},\
  }\href@noop {} {\bibfield  {journal} {\bibinfo  {journal} {Phys. Rev. B}\
  }\textbf {\bibinfo {volume} {86}},\ \bibinfo {pages} {155121} (\bibinfo
  {year} {2012})}\BibitemShut {NoStop}%
\bibitem [{\citenamefont {Bieder}\ and\ \citenamefont
  {Amadon}(2014)}]{Amadon:13}%
  \BibitemOpen
  \bibfield  {author} {\bibinfo {author} {\bibfnamefont {J.}~\bibnamefont
  {Bieder}}\ and\ \bibinfo {author} {\bibfnamefont {B.}~\bibnamefont
  {Amadon}},\ }\href@noop {} {\bibfield  {journal} {\bibinfo  {journal} {Phys.
  Rev. B}\ }\textbf {\bibinfo {volume} {89}},\ \bibinfo {pages} {195132}
  (\bibinfo {year} {2014})}\BibitemShut {NoStop}%
\bibitem [{\citenamefont {Bl\"ochl}(1994)}]{Blochl:1994}%
  \BibitemOpen
  \bibfield  {author} {\bibinfo {author} {\bibfnamefont {P.~E.}\ \bibnamefont
  {Bl\"ochl}},\ }\href {\doibase 10.1103/PhysRevB.50.17953} {\bibfield
  {journal} {\bibinfo  {journal} {Phys. Rev. B}\ }\textbf {\bibinfo {volume}
  {50}},\ \bibinfo {pages} {17953} (\bibinfo {year} {1994})}\BibitemShut
  {NoStop}%
\bibitem [{\citenamefont {Park}\ \emph
  {et~al.}(2014{\natexlab{a}})\citenamefont {Park}, \citenamefont {Millis},\
  and\ \citenamefont {Marianetti}}]{Park:13}%
  \BibitemOpen
  \bibfield  {author} {\bibinfo {author} {\bibfnamefont {H.}~\bibnamefont
  {Park}}, \bibinfo {author} {\bibfnamefont {A.~J.}\ \bibnamefont {Millis}}, \
  and\ \bibinfo {author} {\bibfnamefont {C.~A.}\ \bibnamefont {Marianetti}},\
  }\href {\doibase 10.1103/PhysRevB.89.245133} {\bibfield  {journal} {\bibinfo
  {journal} {Phys. Rev. B}\ }\textbf {\bibinfo {volume} {89}},\ \bibinfo
  {pages} {245133} (\bibinfo {year} {2014}{\natexlab{a}})}\BibitemShut
  {NoStop}%
\bibitem [{\citenamefont {Marzari}\ and\ \citenamefont
  {Vanderbilt}(1997)}]{Marzari:97}%
  \BibitemOpen
  \bibfield  {author} {\bibinfo {author} {\bibfnamefont {N.}~\bibnamefont
  {Marzari}}\ and\ \bibinfo {author} {\bibfnamefont {D.}~\bibnamefont
  {Vanderbilt}},\ }\href {\doibase 10.1103/PhysRevB.56.12847} {\bibfield
  {journal} {\bibinfo  {journal} {Phys. Rev. B}\ }\textbf {\bibinfo {volume}
  {56}},\ \bibinfo {pages} {12847} (\bibinfo {year} {1997})}\BibitemShut
  {NoStop}%
\bibitem [{\citenamefont {Marzari}\ \emph {et~al.}(2012)\citenamefont
  {Marzari}, \citenamefont {Mostofi}, \citenamefont {Yates}, \citenamefont
  {Souza},\ and\ \citenamefont {Vanderbilt}}]{Marzari:12}%
  \BibitemOpen
  \bibfield  {author} {\bibinfo {author} {\bibfnamefont {N.}~\bibnamefont
  {Marzari}}, \bibinfo {author} {\bibfnamefont {A.~A.}\ \bibnamefont
  {Mostofi}}, \bibinfo {author} {\bibfnamefont {J.~R.}\ \bibnamefont {Yates}},
  \bibinfo {author} {\bibfnamefont {I.}~\bibnamefont {Souza}}, \ and\ \bibinfo
  {author} {\bibfnamefont {D.}~\bibnamefont {Vanderbilt}},\ }\href {\doibase
  10.1103/RevModPhys.84.1419} {\bibfield  {journal} {\bibinfo  {journal} {Rev.
  Mod. Phys.}\ }\textbf {\bibinfo {volume} {84}},\ \bibinfo {pages} {1419}
  (\bibinfo {year} {2012})}\BibitemShut {NoStop}%
\bibitem [{\citenamefont {Mostofi}\ \emph {et~al.}(2008)\citenamefont
  {Mostofi}, \citenamefont {Yates}, \citenamefont {Lee}, \citenamefont {Souza},
  \citenamefont {Vanderbilt},\ and\ \citenamefont {Marzari}}]{Wannier}%
  \BibitemOpen
  \bibfield  {author} {\bibinfo {author} {\bibfnamefont {A.~A.}\ \bibnamefont
  {Mostofi}}, \bibinfo {author} {\bibfnamefont {J.~R.}\ \bibnamefont {Yates}},
  \bibinfo {author} {\bibfnamefont {Y.-S.}\ \bibnamefont {Lee}}, \bibinfo
  {author} {\bibfnamefont {I.}~\bibnamefont {Souza}}, \bibinfo {author}
  {\bibfnamefont {D.}~\bibnamefont {Vanderbilt}}, \ and\ \bibinfo {author}
  {\bibfnamefont {N.}~\bibnamefont {Marzari}},\ }\href@noop {} {\bibfield
  {journal} {\bibinfo  {journal} {Computer Physics Communications}\ }\textbf
  {\bibinfo {volume} {178}},\ \bibinfo {pages} {685} (\bibinfo {year}
  {2008})}\BibitemShut {NoStop}%
\bibitem [{\citenamefont {Anisimov}\ \emph {et~al.}(1991)\citenamefont
  {Anisimov}, \citenamefont {Zaanen},\ and\ \citenamefont
  {Andersen}}]{Anisimov:91}%
  \BibitemOpen
  \bibfield  {author} {\bibinfo {author} {\bibfnamefont {V.~I.}\ \bibnamefont
  {Anisimov}}, \bibinfo {author} {\bibfnamefont {J.}~\bibnamefont {Zaanen}}, \
  and\ \bibinfo {author} {\bibfnamefont {O.~K.}\ \bibnamefont {Andersen}},\
  }\href {\doibase 10.1103/PhysRevB.44.943} {\bibfield  {journal} {\bibinfo
  {journal} {Phys. Rev. B}\ }\textbf {\bibinfo {volume} {44}},\ \bibinfo
  {pages} {943} (\bibinfo {year} {1991})}\BibitemShut {NoStop}%
\bibitem [{\citenamefont {Savrasov}\ and\ \citenamefont
  {Kotliar}(2004)}]{Savrasov:04}%
  \BibitemOpen
  \bibfield  {author} {\bibinfo {author} {\bibfnamefont {S.~Y.}\ \bibnamefont
  {Savrasov}}\ and\ \bibinfo {author} {\bibfnamefont {G.}~\bibnamefont
  {Kotliar}},\ }\href {\doibase 10.1103/PhysRevB.69.245101} {\bibfield
  {journal} {\bibinfo  {journal} {Phys. Rev. B}\ }\textbf {\bibinfo {volume}
  {69}},\ \bibinfo {pages} {245101} (\bibinfo {year} {2004})}\BibitemShut
  {NoStop}%
\bibitem [{\citenamefont {Perdew}\ \emph {et~al.}(1996)\citenamefont {Perdew},
  \citenamefont {Burke},\ and\ \citenamefont {Ernzerhof}}]{Perdew:96}%
  \BibitemOpen
  \bibfield  {author} {\bibinfo {author} {\bibfnamefont {J.~P.}\ \bibnamefont
  {Perdew}}, \bibinfo {author} {\bibfnamefont {K.}~\bibnamefont {Burke}}, \
  and\ \bibinfo {author} {\bibfnamefont {M.}~\bibnamefont {Ernzerhof}},\ }\href
  {\doibase 10.1103/PhysRevLett.77.3865} {\bibfield  {journal} {\bibinfo
  {journal} {Phys. Rev. Lett.}\ }\textbf {\bibinfo {volume} {77}},\ \bibinfo
  {pages} {3865} (\bibinfo {year} {1996})}\BibitemShut {NoStop}%
\bibitem [{\citenamefont {Lechermann}\ \emph {et~al.}(2006)\citenamefont
  {Lechermann}, \citenamefont {Georges}, \citenamefont {Poteryaev},
  \citenamefont {Biermann}, \citenamefont {Posternak}, \citenamefont
  {Yamasaki},\ and\ \citenamefont {Andersen}}]{Lechermann:06}%
  \BibitemOpen
  \bibfield  {author} {\bibinfo {author} {\bibfnamefont {F.}~\bibnamefont
  {Lechermann}}, \bibinfo {author} {\bibfnamefont {A.}~\bibnamefont {Georges}},
  \bibinfo {author} {\bibfnamefont {A.}~\bibnamefont {Poteryaev}}, \bibinfo
  {author} {\bibfnamefont {S.}~\bibnamefont {Biermann}}, \bibinfo {author}
  {\bibfnamefont {M.}~\bibnamefont {Posternak}}, \bibinfo {author}
  {\bibfnamefont {A.}~\bibnamefont {Yamasaki}}, \ and\ \bibinfo {author}
  {\bibfnamefont {O.~K.}\ \bibnamefont {Andersen}},\ }\href@noop {} {\bibfield
  {journal} {\bibinfo  {journal} {Phys. Rev. B}\ }\textbf {\bibinfo {volume}
  {74}},\ \bibinfo {pages} {125120} (\bibinfo {year} {2006})}\BibitemShut
  {NoStop}%
\bibitem [{\citenamefont {Kresse}\ and\ \citenamefont
  {Joubert}(1999)}]{Kresse19991758}%
  \BibitemOpen
  \bibfield  {author} {\bibinfo {author} {\bibfnamefont {G.}~\bibnamefont
  {Kresse}}\ and\ \bibinfo {author} {\bibfnamefont {D.}~\bibnamefont
  {Joubert}},\ }\href@noop {} {\bibfield  {journal} {\bibinfo  {journal} {Phys.
  Rev. B}\ }\textbf {\bibinfo {volume} {59}},\ \bibinfo {pages} {1758}
  (\bibinfo {year} {1999})}\BibitemShut {NoStop}%
\bibitem [{\citenamefont {Kerker}(1981)}]{Kerker:81}%
  \BibitemOpen
  \bibfield  {author} {\bibinfo {author} {\bibfnamefont {G.~P.}\ \bibnamefont
  {Kerker}},\ }\href {\doibase 10.1103/PhysRevB.23.3082} {\bibfield  {journal}
  {\bibinfo  {journal} {Phys. Rev. B}\ }\textbf {\bibinfo {volume} {23}},\
  \bibinfo {pages} {3082} (\bibinfo {year} {1981})}\BibitemShut {NoStop}%
\bibitem [{\citenamefont {Wang}\ \emph {et~al.}(2011)\citenamefont {Wang},
  \citenamefont {Dang},\ and\ \citenamefont {Millis}}]{Wang:11}%
  \BibitemOpen
  \bibfield  {author} {\bibinfo {author} {\bibfnamefont {X.}~\bibnamefont
  {Wang}}, \bibinfo {author} {\bibfnamefont {H.~T.}\ \bibnamefont {Dang}}, \
  and\ \bibinfo {author} {\bibfnamefont {A.~J.}\ \bibnamefont {Millis}},\
  }\href {\doibase 10.1103/PhysRevB.84.073104} {\bibfield  {journal} {\bibinfo
  {journal} {Phys. Rev. B}\ }\textbf {\bibinfo {volume} {84}},\ \bibinfo
  {pages} {073104} (\bibinfo {year} {2011})}\BibitemShut {NoStop}%
\bibitem [{\citenamefont {Wang}\ \emph {et~al.}(2012)\citenamefont {Wang},
  \citenamefont {Han}, \citenamefont {de' Medici}, \citenamefont {Park},
  \citenamefont {Marianetti},\ and\ \citenamefont {Millis}}]{Wang:12}%
  \BibitemOpen
  \bibfield  {author} {\bibinfo {author} {\bibfnamefont {X.}~\bibnamefont
  {Wang}}, \bibinfo {author} {\bibfnamefont {M.~J.}\ \bibnamefont {Han}},
  \bibinfo {author} {\bibfnamefont {L.}~\bibnamefont {de' Medici}}, \bibinfo
  {author} {\bibfnamefont {H.}~\bibnamefont {Park}}, \bibinfo {author}
  {\bibfnamefont {C.~A.}\ \bibnamefont {Marianetti}}, \ and\ \bibinfo {author}
  {\bibfnamefont {A.~J.}\ \bibnamefont {Millis}},\ }\href {\doibase
  10.1103/PhysRevB.86.195136} {\bibfield  {journal} {\bibinfo  {journal} {Phys.
  Rev. B}\ }\textbf {\bibinfo {volume} {86}},\ \bibinfo {pages} {195136}
  (\bibinfo {year} {2012})}\BibitemShut {NoStop}%
\bibitem [{\citenamefont {Park}\ \emph {et~al.}(2012)\citenamefont {Park},
  \citenamefont {Millis},\ and\ \citenamefont {Marianetti}}]{Park:12}%
  \BibitemOpen
  \bibfield  {author} {\bibinfo {author} {\bibfnamefont {H.}~\bibnamefont
  {Park}}, \bibinfo {author} {\bibfnamefont {A.~J.}\ \bibnamefont {Millis}}, \
  and\ \bibinfo {author} {\bibfnamefont {C.~A.}\ \bibnamefont {Marianetti}},\
  }\href {\doibase 10.1103/PhysRevLett.109.156402} {\bibfield  {journal}
  {\bibinfo  {journal} {Phys. Rev. Lett.}\ }\textbf {\bibinfo {volume} {109}},\
  \bibinfo {pages} {156402} (\bibinfo {year} {2012})}\BibitemShut {NoStop}%
\bibitem [{\citenamefont {Dang}\ \emph
  {et~al.}(2014{\natexlab{a}})\citenamefont {Dang}, \citenamefont {Millis},\
  and\ \citenamefont {Marianetti}}]{Dang:13}%
  \BibitemOpen
  \bibfield  {author} {\bibinfo {author} {\bibfnamefont {H.~T.}\ \bibnamefont
  {Dang}}, \bibinfo {author} {\bibfnamefont {A.~J.}\ \bibnamefont {Millis}}, \
  and\ \bibinfo {author} {\bibfnamefont {C.~A.}\ \bibnamefont {Marianetti}},\
  }\href@noop {} {\bibfield  {journal} {\bibinfo  {journal} {Phys. Rev. B}\
  }\textbf {\bibinfo {volume} {89}},\ \bibinfo {pages} {161113} (\bibinfo
  {year} {2014}{\natexlab{a}})}\BibitemShut {NoStop}%
\bibitem [{\citenamefont {Czyzyk}\ and\ \citenamefont
  {Sawatzky}(1994)}]{Sawatzky:94}%
  \BibitemOpen
  \bibfield  {author} {\bibinfo {author} {\bibfnamefont {M.~T.}\ \bibnamefont
  {Czyzyk}}\ and\ \bibinfo {author} {\bibfnamefont {G.~A.}\ \bibnamefont
  {Sawatzky}},\ }\href {\doibase 10.1103/PhysRevB.49.14211} {\bibfield
  {journal} {\bibinfo  {journal} {Phys. Rev. B}\ }\textbf {\bibinfo {volume}
  {49}},\ \bibinfo {pages} {14211} (\bibinfo {year} {1994})}\BibitemShut
  {NoStop}%
\bibitem [{\citenamefont {Amadon}\ \emph {et~al.}(2008)\citenamefont {Amadon},
  \citenamefont {Lechermann}, \citenamefont {Georges}, \citenamefont {Jollet},
  \citenamefont {Wehling},\ and\ \citenamefont {Lichtenstein}}]{Amadon:08}%
  \BibitemOpen
  \bibfield  {author} {\bibinfo {author} {\bibfnamefont {B.}~\bibnamefont
  {Amadon}}, \bibinfo {author} {\bibfnamefont {F.}~\bibnamefont {Lechermann}},
  \bibinfo {author} {\bibfnamefont {A.}~\bibnamefont {Georges}}, \bibinfo
  {author} {\bibfnamefont {F.}~\bibnamefont {Jollet}}, \bibinfo {author}
  {\bibfnamefont {T.~O.}\ \bibnamefont {Wehling}}, \ and\ \bibinfo {author}
  {\bibfnamefont {A.~I.}\ \bibnamefont {Lichtenstein}},\ }\href {\doibase
  10.1103/PhysRevB.77.205112} {\bibfield  {journal} {\bibinfo  {journal} {Phys.
  Rev. B}\ }\textbf {\bibinfo {volume} {77}},\ \bibinfo {pages} {205112}
  (\bibinfo {year} {2008})}\BibitemShut {NoStop}%
\bibitem [{\citenamefont {Karolak}\ \emph {et~al.}(2010)\citenamefont
  {Karolak}, \citenamefont {Ulm}, \citenamefont {Wehling}, \citenamefont
  {Mazurenko}, \citenamefont {Poteryaev},\ and\ \citenamefont
  {Lichtenstein}}]{Karolak:10}%
  \BibitemOpen
  \bibfield  {author} {\bibinfo {author} {\bibfnamefont {M.}~\bibnamefont
  {Karolak}}, \bibinfo {author} {\bibfnamefont {G.}~\bibnamefont {Ulm}},
  \bibinfo {author} {\bibfnamefont {T.}~\bibnamefont {Wehling}}, \bibinfo
  {author} {\bibfnamefont {V.}~\bibnamefont {Mazurenko}}, \bibinfo {author}
  {\bibfnamefont {A.}~\bibnamefont {Poteryaev}}, \ and\ \bibinfo {author}
  {\bibfnamefont {A.}~\bibnamefont {Lichtenstein}},\ }\href@noop {} {\bibfield
  {journal} {\bibinfo  {journal} {Journal of Electron Spectroscopy and Related
  Phenomena}\ }\textbf {\bibinfo {volume} {181}},\ \bibinfo {pages} {11}
  (\bibinfo {year} {2010})}\BibitemShut {NoStop}%
\bibitem [{\citenamefont {Haule}\ \emph {et~al.}(2013)\citenamefont {Haule},
  \citenamefont {Birol},\ and\ \citenamefont {Kotliar}}]{Haule:13}%
  \BibitemOpen
  \bibfield  {author} {\bibinfo {author} {\bibfnamefont {K.}~\bibnamefont
  {Haule}}, \bibinfo {author} {\bibfnamefont {T.}~\bibnamefont {Birol}}, \ and\
  \bibinfo {author} {\bibfnamefont {G.}~\bibnamefont {Kotliar}},\ }\href@noop
  {} {\bibfield  {journal} {\bibinfo  {journal} {arXiv}\ ,\ \bibinfo {pages}
  {9}} (\bibinfo {year} {2013})},\ \Eprint {http://arxiv.org/abs/1310.1158}
  {1310.1158} \BibitemShut {NoStop}%
\bibitem [{\citenamefont {Anisimov}\ \emph {et~al.}(1993)\citenamefont
  {Anisimov}, \citenamefont {Solovyev}, \citenamefont {Korotin}, \citenamefont
  {Czyzyk},\ and\ \citenamefont {Sawatzky}}]{Anisimov:93}%
  \BibitemOpen
  \bibfield  {author} {\bibinfo {author} {\bibfnamefont {V.~I.}\ \bibnamefont
  {Anisimov}}, \bibinfo {author} {\bibfnamefont {I.~V.}\ \bibnamefont
  {Solovyev}}, \bibinfo {author} {\bibfnamefont {M.~A.}\ \bibnamefont
  {Korotin}}, \bibinfo {author} {\bibfnamefont {M.~T.}\ \bibnamefont {Czyzyk}},
  \ and\ \bibinfo {author} {\bibfnamefont {G.~A.}\ \bibnamefont {Sawatzky}},\
  }\href {\doibase 10.1103/PhysRevB.48.16929} {\bibfield  {journal} {\bibinfo
  {journal} {Phys. Rev. B}\ }\textbf {\bibinfo {volume} {48}},\ \bibinfo
  {pages} {16929} (\bibinfo {year} {1993})}\BibitemShut {NoStop}%
\bibitem [{\citenamefont {Petukhov}\ \emph {et~al.}(2003)\citenamefont
  {Petukhov}, \citenamefont {Mazin}, \citenamefont {Chioncel},\ and\
  \citenamefont {Lichtenstein}}]{Petukhov:03}%
  \BibitemOpen
  \bibfield  {author} {\bibinfo {author} {\bibfnamefont {A.~G.}\ \bibnamefont
  {Petukhov}}, \bibinfo {author} {\bibfnamefont {I.~I.}\ \bibnamefont {Mazin}},
  \bibinfo {author} {\bibfnamefont {L.}~\bibnamefont {Chioncel}}, \ and\
  \bibinfo {author} {\bibfnamefont {A.~I.}\ \bibnamefont {Lichtenstein}},\
  }\href {\doibase 10.1103/PhysRevB.67.153106} {\bibfield  {journal} {\bibinfo
  {journal} {Phys. Rev. B}\ }\textbf {\bibinfo {volume} {67}},\ \bibinfo
  {pages} {153106} (\bibinfo {year} {2003})}\BibitemShut {NoStop}%
\bibitem [{\citenamefont {Haule}\ \emph {et~al.}(2010)\citenamefont {Haule},
  \citenamefont {Yee},\ and\ \citenamefont {Kim}}]{Haule:10}%
  \BibitemOpen
  \bibfield  {author} {\bibinfo {author} {\bibfnamefont {K.}~\bibnamefont
  {Haule}}, \bibinfo {author} {\bibfnamefont {C.-H.}\ \bibnamefont {Yee}}, \
  and\ \bibinfo {author} {\bibfnamefont {K.}~\bibnamefont {Kim}},\ }\href
  {\doibase 10.1103/PhysRevB.81.195107} {\bibfield  {journal} {\bibinfo
  {journal} {Phys. Rev. B}\ }\textbf {\bibinfo {volume} {81}},\ \bibinfo
  {pages} {195107} (\bibinfo {year} {2010})}\BibitemShut {NoStop}%
\bibitem [{\citenamefont {Medarde}\ \emph {et~al.}(1997)\citenamefont
  {Medarde}, \citenamefont {Mesot}, \citenamefont {Rosenkranz}, \citenamefont
  {Lacorre}, \citenamefont {Marshall}, \citenamefont {Klotz}, \citenamefont
  {Loveday}, \citenamefont {Hamel}, \citenamefont {Hull},\ and\ \citenamefont
  {Radaelli}}]{Medarde:1997}%
  \BibitemOpen
  \bibfield  {author} {\bibinfo {author} {\bibfnamefont {M.}~\bibnamefont
  {Medarde}}, \bibinfo {author} {\bibfnamefont {J.}~\bibnamefont {Mesot}},
  \bibinfo {author} {\bibfnamefont {S.}~\bibnamefont {Rosenkranz}}, \bibinfo
  {author} {\bibfnamefont {P.}~\bibnamefont {Lacorre}}, \bibinfo {author}
  {\bibfnamefont {W.}~\bibnamefont {Marshall}}, \bibinfo {author}
  {\bibfnamefont {S.}~\bibnamefont {Klotz}}, \bibinfo {author} {\bibfnamefont
  {J.}~\bibnamefont {Loveday}}, \bibinfo {author} {\bibfnamefont
  {G.}~\bibnamefont {Hamel}}, \bibinfo {author} {\bibfnamefont
  {S.}~\bibnamefont {Hull}}, \ and\ \bibinfo {author} {\bibfnamefont
  {P.}~\bibnamefont {Radaelli}},\ }\href@noop {} {\bibfield  {journal}
  {\bibinfo  {journal} {Physica B: Condensed Matter}\ }\textbf {\bibinfo
  {volume} {234-236}},\ \bibinfo {pages} {15 } (\bibinfo {year}
  {1997})}\BibitemShut {NoStop}%
\bibitem [{\citenamefont {Zhou}\ \emph {et~al.}(2004)\citenamefont {Zhou},
  \citenamefont {Goodenough},\ and\ \citenamefont {Dabrowski}}]{Zhou:04}%
  \BibitemOpen
  \bibfield  {author} {\bibinfo {author} {\bibfnamefont {J.-S.}\ \bibnamefont
  {Zhou}}, \bibinfo {author} {\bibfnamefont {J.~B.}\ \bibnamefont
  {Goodenough}}, \ and\ \bibinfo {author} {\bibfnamefont {B.}~\bibnamefont
  {Dabrowski}},\ }\href {\doibase 10.1103/PhysRevB.70.081102} {\bibfield
  {journal} {\bibinfo  {journal} {Phys. Rev. B}\ }\textbf {\bibinfo {volume}
  {70}},\ \bibinfo {pages} {081102} (\bibinfo {year} {2004})}\BibitemShut
  {NoStop}%
\bibitem [{\citenamefont {Amboage}\ \emph {et~al.}(2005)\citenamefont
  {Amboage}, \citenamefont {Hanfland}, \citenamefont {Alonso},\ and\
  \citenamefont {Martínez-Lope}}]{Amboage:05}%
  \BibitemOpen
  \bibfield  {author} {\bibinfo {author} {\bibfnamefont {M.}~\bibnamefont
  {Amboage}}, \bibinfo {author} {\bibfnamefont {M.}~\bibnamefont {Hanfland}},
  \bibinfo {author} {\bibfnamefont {J.~A.}\ \bibnamefont {Alonso}}, \ and\
  \bibinfo {author} {\bibfnamefont {M.~J.}\ \bibnamefont {Martínez-Lope}},\
  }\href@noop {} {\bibfield  {journal} {\bibinfo  {journal} {Journal of
  Physics: Condensed Matter}\ }\textbf {\bibinfo {volume} {17}},\ \bibinfo
  {pages} {S783} (\bibinfo {year} {2005})}\BibitemShut {NoStop}%
\bibitem [{\citenamefont {Garcia-Mu\~noz}\ \emph {et~al.}(2004)\citenamefont
  {Garcia-Mu\~noz}, \citenamefont {Amboage}, \citenamefont {Hanfland},
  \citenamefont {Alonso}, \citenamefont {Martinez-Lope},\ and\ \citenamefont
  {Mortimer}}]{Amboage:04}%
  \BibitemOpen
  \bibfield  {author} {\bibinfo {author} {\bibfnamefont {J.~L.}\ \bibnamefont
  {Garcia-Mu\~noz}}, \bibinfo {author} {\bibfnamefont {M.}~\bibnamefont
  {Amboage}}, \bibinfo {author} {\bibfnamefont {M.}~\bibnamefont {Hanfland}},
  \bibinfo {author} {\bibfnamefont {J.~A.}\ \bibnamefont {Alonso}}, \bibinfo
  {author} {\bibfnamefont {M.~J.}\ \bibnamefont {Martinez-Lope}}, \ and\
  \bibinfo {author} {\bibfnamefont {R.}~\bibnamefont {Mortimer}},\ }\href
  {\doibase 10.1103/PhysRevB.69.094106} {\bibfield  {journal} {\bibinfo
  {journal} {Phys. Rev. B}\ }\textbf {\bibinfo {volume} {69}},\ \bibinfo
  {pages} {094106} (\bibinfo {year} {2004})}\BibitemShut {NoStop}%
\bibitem [{\citenamefont {Ramos}\ \emph {et~al.}(2012)\citenamefont {Ramos},
  \citenamefont {Piamonteze}, \citenamefont {Tolentino}, \citenamefont
  {Souza-Neto}, \citenamefont {Bunau}, \citenamefont {Joly}, \citenamefont
  {Grenier}, \citenamefont {Iti\'e}, \citenamefont {Massa}, \citenamefont
  {Alonso},\ and\ \citenamefont {Martinez-Lope}}]{Ramos:12}%
  \BibitemOpen
  \bibfield  {author} {\bibinfo {author} {\bibfnamefont {A.~Y.}\ \bibnamefont
  {Ramos}}, \bibinfo {author} {\bibfnamefont {C.}~\bibnamefont {Piamonteze}},
  \bibinfo {author} {\bibfnamefont {H.~C.~N.}\ \bibnamefont {Tolentino}},
  \bibinfo {author} {\bibfnamefont {N.~M.}\ \bibnamefont {Souza-Neto}},
  \bibinfo {author} {\bibfnamefont {O.}~\bibnamefont {Bunau}}, \bibinfo
  {author} {\bibfnamefont {Y.}~\bibnamefont {Joly}}, \bibinfo {author}
  {\bibfnamefont {S.}~\bibnamefont {Grenier}}, \bibinfo {author} {\bibfnamefont
  {J.-P.}\ \bibnamefont {Iti\'e}}, \bibinfo {author} {\bibfnamefont {N.~E.}\
  \bibnamefont {Massa}}, \bibinfo {author} {\bibfnamefont {J.~A.}\ \bibnamefont
  {Alonso}}, \ and\ \bibinfo {author} {\bibfnamefont {M.~J.}\ \bibnamefont
  {Martinez-Lope}},\ }\href@noop {} {\bibfield  {journal} {\bibinfo  {journal}
  {Phys. Rev. B}\ }\textbf {\bibinfo {volume} {85}},\ \bibinfo {pages} {045102}
  (\bibinfo {year} {2012})}\BibitemShut {NoStop}%
\bibitem [{\citenamefont {Medarde}(1997)}]{Medarde:97}%
  \BibitemOpen
  \bibfield  {author} {\bibinfo {author} {\bibfnamefont {M.~L.}\ \bibnamefont
  {Medarde}},\ }\href@noop {} {\bibfield  {journal} {\bibinfo  {journal}
  {Journal of Physics: Condensed Matter}\ }\textbf {\bibinfo {volume} {9}},\
  \bibinfo {pages} {1679} (\bibinfo {year} {1997})}\BibitemShut {NoStop}%
\bibitem [{\citenamefont {Kresse}\ and\ \citenamefont
  {Hafner}(1994)}]{PhysRevB.49.14251}%
  \BibitemOpen
  \bibfield  {author} {\bibinfo {author} {\bibfnamefont {G.}~\bibnamefont
  {Kresse}}\ and\ \bibinfo {author} {\bibfnamefont {J.}~\bibnamefont
  {Hafner}},\ }\href {\doibase 10.1103/PhysRevB.49.14251} {\bibfield  {journal}
  {\bibinfo  {journal} {Phys. Rev. B}\ }\textbf {\bibinfo {volume} {49}},\
  \bibinfo {pages} {14251} (\bibinfo {year} {1994})}\BibitemShut {NoStop}%
\bibitem [{\citenamefont {Kresse}\ and\ \citenamefont
  {Hafner}(1993)}]{PhysRevB.47.558}%
  \BibitemOpen
  \bibfield  {author} {\bibinfo {author} {\bibfnamefont {G.}~\bibnamefont
  {Kresse}}\ and\ \bibinfo {author} {\bibfnamefont {J.}~\bibnamefont
  {Hafner}},\ }\href {\doibase 10.1103/PhysRevB.47.558} {\bibfield  {journal}
  {\bibinfo  {journal} {Phys. Rev. B}\ }\textbf {\bibinfo {volume} {47}},\
  \bibinfo {pages} {558} (\bibinfo {year} {1993})}\BibitemShut {NoStop}%
\bibitem [{\citenamefont {Kresse}\ and\ \citenamefont
  {Furthmüller}(1996)}]{Kresse199615}%
  \BibitemOpen
  \bibfield  {author} {\bibinfo {author} {\bibfnamefont {G.}~\bibnamefont
  {Kresse}}\ and\ \bibinfo {author} {\bibfnamefont {J.}~\bibnamefont
  {Furthmüller}},\ }\href@noop {} {\bibfield  {journal} {\bibinfo  {journal}
  {Computational Materials Science}\ }\textbf {\bibinfo {volume} {6}},\
  \bibinfo {pages} {15 } (\bibinfo {year} {1996})}\BibitemShut {NoStop}%
\bibitem [{\citenamefont {Kresse}\ and\ \citenamefont
  {Furthmuller}(1996)}]{Kresse199611169}%
  \BibitemOpen
  \bibfield  {author} {\bibinfo {author} {\bibfnamefont {G.}~\bibnamefont
  {Kresse}}\ and\ \bibinfo {author} {\bibfnamefont {J.}~\bibnamefont
  {Furthmuller}},\ }\href@noop {} {\bibfield  {journal} {\bibinfo  {journal}
  {Phys. Rev. B}\ }\textbf {\bibinfo {volume} {54}},\ \bibinfo {pages} {11169}
  (\bibinfo {year} {1996})}\BibitemShut {NoStop}%
\bibitem [{\citenamefont {Georges}\ \emph {et~al.}(1996)\citenamefont
  {Georges}, \citenamefont {Kotliar}, \citenamefont {Krauth},\ and\
  \citenamefont {Rozenberg}}]{Georges:96}%
  \BibitemOpen
  \bibfield  {author} {\bibinfo {author} {\bibfnamefont {A.}~\bibnamefont
  {Georges}}, \bibinfo {author} {\bibfnamefont {G.}~\bibnamefont {Kotliar}},
  \bibinfo {author} {\bibfnamefont {W.}~\bibnamefont {Krauth}}, \ and\ \bibinfo
  {author} {\bibfnamefont {M.~J.}\ \bibnamefont {Rozenberg}},\ }\href {\doibase
  10.1103/RevModPhys.68.13} {\bibfield  {journal} {\bibinfo  {journal} {Rev.
  Mod. Phys.}\ }\textbf {\bibinfo {volume} {68}},\ \bibinfo {pages} {13}
  (\bibinfo {year} {1996})}\BibitemShut {NoStop}%
\bibitem [{\citenamefont {Werner}\ \emph {et~al.}(2009)\citenamefont {Werner},
  \citenamefont {Gull}, \citenamefont {Parcollet},\ and\ \citenamefont
  {Millis}}]{Werner098site}%
  \BibitemOpen
  \bibfield  {author} {\bibinfo {author} {\bibfnamefont {P.}~\bibnamefont
  {Werner}}, \bibinfo {author} {\bibfnamefont {E.}~\bibnamefont {Gull}},
  \bibinfo {author} {\bibfnamefont {O.}~\bibnamefont {Parcollet}}, \ and\
  \bibinfo {author} {\bibfnamefont {A.~J.}\ \bibnamefont {Millis}},\ }\href
  {\doibase 10.1103/PhysRevB.80.045120} {\bibfield  {journal} {\bibinfo
  {journal} {Phys. Rev. B}\ }\textbf {\bibinfo {volume} {80}},\ \bibinfo {eid}
  {045120} (\bibinfo {year} {2009})}\BibitemShut {NoStop}%
\bibitem [{\citenamefont {Cheng}\ \emph {et~al.}(2010)\citenamefont {Cheng},
  \citenamefont {Zhou}, \citenamefont {Goodenough}, \citenamefont {Alonso},\
  and\ \citenamefont {Martinez-Lope}}]{Cheng:10}%
  \BibitemOpen
  \bibfield  {author} {\bibinfo {author} {\bibfnamefont {J.-G.}\ \bibnamefont
  {Cheng}}, \bibinfo {author} {\bibfnamefont {J.-S.}\ \bibnamefont {Zhou}},
  \bibinfo {author} {\bibfnamefont {J.~B.}\ \bibnamefont {Goodenough}},
  \bibinfo {author} {\bibfnamefont {J.~A.}\ \bibnamefont {Alonso}}, \ and\
  \bibinfo {author} {\bibfnamefont {M.~J.}\ \bibnamefont {Martinez-Lope}},\
  }\href {\doibase 10.1103/PhysRevB.82.085107} {\bibfield  {journal} {\bibinfo
  {journal} {Phys. Rev. B}\ }\textbf {\bibinfo {volume} {82}},\ \bibinfo
  {pages} {085107} (\bibinfo {year} {2010})}\BibitemShut {NoStop}%
\bibitem [{\citenamefont {Park}\ \emph
  {et~al.}(2014{\natexlab{b}})\citenamefont {Park}, \citenamefont {Millis},\
  and\ \citenamefont {Marianetti}}]{Park:14}%
  \BibitemOpen
  \bibfield  {author} {\bibinfo {author} {\bibfnamefont {H.}~\bibnamefont
  {Park}}, \bibinfo {author} {\bibfnamefont {A.~J.}\ \bibnamefont {Millis}}, \
  and\ \bibinfo {author} {\bibfnamefont {C.~A.}\ \bibnamefont {Marianetti}},\
  }\href@noop {} {\bibfield  {journal} {\bibinfo  {journal} {To be published}\
  } (\bibinfo {year} {2014}{\natexlab{b}})}\BibitemShut {NoStop}%
\bibitem [{\citenamefont {Alonso}\ \emph {et~al.}(2001)\citenamefont {Alonso},
  \citenamefont {Martinez-Lope}, \citenamefont {Casais}, \citenamefont
  {Garcia-Munoz}, \citenamefont {Fernandez-Diaz},\ and\ \citenamefont
  {Aranda}}]{Alonso:01}%
  \BibitemOpen
  \bibfield  {author} {\bibinfo {author} {\bibfnamefont {J.~A.}\ \bibnamefont
  {Alonso}}, \bibinfo {author} {\bibfnamefont {M.~J.}\ \bibnamefont
  {Martinez-Lope}}, \bibinfo {author} {\bibfnamefont {M.~T.}\ \bibnamefont
  {Casais}}, \bibinfo {author} {\bibfnamefont {J.~L.}\ \bibnamefont
  {Garcia-Munoz}}, \bibinfo {author} {\bibfnamefont {M.~T.}\ \bibnamefont
  {Fernandez-Diaz}}, \ and\ \bibinfo {author} {\bibfnamefont {M.~A.~G.}\
  \bibnamefont {Aranda}},\ }\href@noop {} {\bibfield  {journal} {\bibinfo
  {journal} {Phys. Rev. B}\ }\textbf {\bibinfo {volume} {64}},\ \bibinfo
  {pages} {094102} (\bibinfo {year} {2001})}\BibitemShut {NoStop}%
\bibitem [{\citenamefont {Garcia-Mu\~noz}\ \emph {et~al.}(1992)\citenamefont
  {Garcia-Mu\~noz}, \citenamefont {Rodriguez-Carvajal}, \citenamefont
  {Lacorre},\ and\ \citenamefont {Torrance}}]{Garcia:92}%
  \BibitemOpen
  \bibfield  {author} {\bibinfo {author} {\bibfnamefont {J.~L.}\ \bibnamefont
  {Garcia-Mu\~noz}}, \bibinfo {author} {\bibfnamefont {J.}~\bibnamefont
  {Rodriguez-Carvajal}}, \bibinfo {author} {\bibfnamefont {P.}~\bibnamefont
  {Lacorre}}, \ and\ \bibinfo {author} {\bibfnamefont {J.~B.}\ \bibnamefont
  {Torrance}},\ }\href {\doibase 10.1103/PhysRevB.46.4414} {\bibfield
  {journal} {\bibinfo  {journal} {Phys. Rev. B}\ }\textbf {\bibinfo {volume}
  {46}},\ \bibinfo {pages} {4414} (\bibinfo {year} {1992})}\BibitemShut
  {NoStop}%
\bibitem [{\citenamefont {Garcia-Mu\~noz}\ \emph {et~al.}(2009)\citenamefont
  {Garcia-Mu\~noz}, \citenamefont {Aranda}, \citenamefont {Alonso},\ and\
  \citenamefont {Martinez-Lope}}]{Garcia:09}%
  \BibitemOpen
  \bibfield  {author} {\bibinfo {author} {\bibfnamefont {J.~L.}\ \bibnamefont
  {Garcia-Mu\~noz}}, \bibinfo {author} {\bibfnamefont {M.~A.~G.}\ \bibnamefont
  {Aranda}}, \bibinfo {author} {\bibfnamefont {J.~A.}\ \bibnamefont {Alonso}},
  \ and\ \bibinfo {author} {\bibfnamefont {M.~J.}\ \bibnamefont
  {Martinez-Lope}},\ }\href {\doibase 10.1103/PhysRevB.79.134432} {\bibfield
  {journal} {\bibinfo  {journal} {Phys. Rev. B}\ }\textbf {\bibinfo {volume}
  {79}},\ \bibinfo {pages} {134432} (\bibinfo {year} {2009})}\BibitemShut
  {NoStop}%
\bibitem [{\citenamefont {Dang}\ \emph
  {et~al.}(2014{\natexlab{b}})\citenamefont {Dang}, \citenamefont {Millis},\
  and\ \citenamefont {Marianetti}}]{Dang:14a}%
  \BibitemOpen
  \bibfield  {author} {\bibinfo {author} {\bibfnamefont {H.~T.}\ \bibnamefont
  {Dang}}, \bibinfo {author} {\bibfnamefont {A.~J.}\ \bibnamefont {Millis}}, \
  and\ \bibinfo {author} {\bibfnamefont {C.~A.}\ \bibnamefont {Marianetti}},\
  }\href@noop {} {\bibfield  {journal} {\bibinfo  {journal} {Phys. Rev. B}\
  }\textbf {\bibinfo {volume} {89}},\ \bibinfo {pages} {161113} (\bibinfo
  {year} {2014}{\natexlab{b}})}\BibitemShut {NoStop}%
\bibitem [{\citenamefont {Dang}\ \emph
  {et~al.}(2014{\natexlab{c}})\citenamefont {Dang}, \citenamefont {Ai},
  \citenamefont {Millis},\ and\ \citenamefont {Marianetti}}]{Dang:14b}%
  \BibitemOpen
  \bibfield  {author} {\bibinfo {author} {\bibfnamefont {H.~T.}\ \bibnamefont
  {Dang}}, \bibinfo {author} {\bibfnamefont {X.}~\bibnamefont {Ai}}, \bibinfo
  {author} {\bibfnamefont {A.~J.}\ \bibnamefont {Millis}}, \ and\ \bibinfo
  {author} {\bibfnamefont {C.~A.}\ \bibnamefont {Marianetti}},\ }\href@noop {}
  {\bibfield  {journal} {\bibinfo  {journal} {arXiv:1407.6505}\ } (\bibinfo
  {year} {2014}{\natexlab{c}})}\BibitemShut {NoStop}%
\bibitem [{\citenamefont {Horiba}\ \emph {et~al.}(2007)\citenamefont {Horiba},
  \citenamefont {Eguchi}, \citenamefont {Taguchi}, \citenamefont {Chainani},
  \citenamefont {Kikkawa}, \citenamefont {Senba}, \citenamefont {Ohashi},\ and\
  \citenamefont {Shin}}]{Horiba:07}%
  \BibitemOpen
  \bibfield  {author} {\bibinfo {author} {\bibfnamefont {K.}~\bibnamefont
  {Horiba}}, \bibinfo {author} {\bibfnamefont {R.}~\bibnamefont {Eguchi}},
  \bibinfo {author} {\bibfnamefont {M.}~\bibnamefont {Taguchi}}, \bibinfo
  {author} {\bibfnamefont {A.}~\bibnamefont {Chainani}}, \bibinfo {author}
  {\bibfnamefont {A.}~\bibnamefont {Kikkawa}}, \bibinfo {author} {\bibfnamefont
  {Y.}~\bibnamefont {Senba}}, \bibinfo {author} {\bibfnamefont
  {H.}~\bibnamefont {Ohashi}}, \ and\ \bibinfo {author} {\bibfnamefont
  {S.}~\bibnamefont {Shin}},\ }\href {\doibase 10.1103/PhysRevB.76.155104}
  {\bibfield  {journal} {\bibinfo  {journal} {Phys. Rev. B}\ }\textbf {\bibinfo
  {volume} {76}},\ \bibinfo {pages} {155104} (\bibinfo {year}
  {2007})}\BibitemShut {NoStop}%
\bibitem [{\citenamefont {Biermann}\ \emph {et~al.}(2005)\citenamefont
  {Biermann}, \citenamefont {Poteryaev}, \citenamefont {Lichtenstein},\ and\
  \citenamefont {Georges}}]{Biermann05}%
  \BibitemOpen
  \bibfield  {author} {\bibinfo {author} {\bibfnamefont {S.}~\bibnamefont
  {Biermann}}, \bibinfo {author} {\bibfnamefont {A.}~\bibnamefont {Poteryaev}},
  \bibinfo {author} {\bibfnamefont {A.~I.}\ \bibnamefont {Lichtenstein}}, \
  and\ \bibinfo {author} {\bibfnamefont {A.}~\bibnamefont {Georges}},\ }\href
  {\doibase 10.1103/PhysRevLett.94.026404} {\bibfield  {journal} {\bibinfo
  {journal} {Phys. Rev. Lett.}\ }\textbf {\bibinfo {volume} {94}},\ \bibinfo
  {pages} {026404} (\bibinfo {year} {2005})}\BibitemShut {NoStop}%
\bibitem [{\citenamefont {Leonov}\ \emph {et~al.}(2014)\citenamefont {Leonov},
  \citenamefont {Anisimov},\ and\ \citenamefont {Vollhardt}}]{Leonov:14}%
  \BibitemOpen
  \bibfield  {author} {\bibinfo {author} {\bibfnamefont {I.}~\bibnamefont
  {Leonov}}, \bibinfo {author} {\bibfnamefont {V.~I.}\ \bibnamefont
  {Anisimov}}, \ and\ \bibinfo {author} {\bibfnamefont {D.}~\bibnamefont
  {Vollhardt}},\ }\href {\doibase 10.1103/PhysRevLett.112.146401} {\bibfield
  {journal} {\bibinfo  {journal} {Phys. Rev. Lett.}\ }\textbf {\bibinfo
  {volume} {112}},\ \bibinfo {pages} {146401} (\bibinfo {year}
  {2014})}\BibitemShut {NoStop}%
\bibitem [{\citenamefont {Aryasetiawan}\ \emph {et~al.}(2006)\citenamefont
  {Aryasetiawan}, \citenamefont {Karlsson}, \citenamefont {Jepsen},\ and\
  \citenamefont {Sch\"onberger}}]{Aryasetiawan06}%
  \BibitemOpen
  \bibfield  {author} {\bibinfo {author} {\bibfnamefont {F.}~\bibnamefont
  {Aryasetiawan}}, \bibinfo {author} {\bibfnamefont {K.}~\bibnamefont
  {Karlsson}}, \bibinfo {author} {\bibfnamefont {O.}~\bibnamefont {Jepsen}}, \
  and\ \bibinfo {author} {\bibfnamefont {U.}~\bibnamefont {Sch\"onberger}},\
  }\href {\doibase 10.1103/PhysRevB.74.125106} {\bibfield  {journal} {\bibinfo
  {journal} {Phys. Rev. B}\ }\textbf {\bibinfo {volume} {74}},\ \bibinfo
  {pages} {125106} (\bibinfo {year} {2006})}\BibitemShut {NoStop}%
\bibitem [{\citenamefont {Vaugier}\ \emph {et~al.}(2012)\citenamefont
  {Vaugier}, \citenamefont {Jiang},\ and\ \citenamefont
  {Biermann}}]{Vaugier12}%
  \BibitemOpen
  \bibfield  {author} {\bibinfo {author} {\bibfnamefont {L.}~\bibnamefont
  {Vaugier}}, \bibinfo {author} {\bibfnamefont {H.}~\bibnamefont {Jiang}}, \
  and\ \bibinfo {author} {\bibfnamefont {S.}~\bibnamefont {Biermann}},\ }\href
  {\doibase 10.1103/PhysRevB.86.165105} {\bibfield  {journal} {\bibinfo
  {journal} {Phys. Rev. B}\ }\textbf {\bibinfo {volume} {86}},\ \bibinfo
  {pages} {165105} (\bibinfo {year} {2012})}\BibitemShut {NoStop}%
\end{thebibliography}%

\end{document}